\documentclass[amsmath,amssymb,prd,notitlepage,longbibliography,superscriptaddress,nofootinbib]{revtex4-2}

\usepackage{amsmath}
\usepackage{enumerate}
\usepackage{amsfonts}
\usepackage[utf8]{inputenc}
\usepackage[T1]{fontenc}
\usepackage{hyperref}

\usepackage{bbm}
\usepackage{amssymb}
\usepackage{amsthm}
\usepackage{mathtools}
\usepackage{filecontents}
\usepackage{comment}
\usepackage{mathrsfs}

\makeatletter
\def\@fpheader{\relax}
\makeatother

\newcounter{parentsubequation}% Counter for ``parent equation''.

\makeatother

\DeclareMathAlphabet{\mathbbold}{U}{bbold}{m}{n} 

\DeclareMathOperator{\Tr}{Tr}

\begin{document}

\title{Scalar/Dirac perturbations of the Reissner-Nordström black hole and Painlevé
transcendents}

\author{João Paulo Cavalcante,}
\email{joaopaulocavalcante@hotmail.com.br}
\author{Bruno Carneiro da Cunha}
\email{bruno.ccunha@ufpe.br}
\affiliation{Departamento de Física, Universidade Federal de
  Pernambuco, 50670-901, Recife, Brazil}

\begin{abstract}
We investigate spin-$0$ and spin-$\frac{1}{2}$ perturbations for
non-extremal and extremal Reissner-Nordström backgrounds using the
isomonodromic method. We calculate the fundamental quasinormal modes
(QNMs) associated with each perturbation as a function of the
electromagnetic coupling $qQ$ and ratio $Q/M$. After corroborating the
literature values for generic $qQ$ and $Q/M$, we turn to study the
near-extremal limit. In parallel with the study of QNMs for the Kerr
geometry, we find the existence of ``non-damping'' modes for $qQ$
above a spin-dependent critical value, and locate the bifurcation
point in the $q-Q$ parameter space. 
\end{abstract}


\maketitle

\section{Introduction}
\label{intr}

The Reissner-Nordström (RN) metric is a solution to the
Einstein-Maxwell field equations, corresponding to a non-rotating
charged black hole of mass $M$ and charge $Q$. In static and
spherically symmetric coordinates, the metric is given by \cite{Wald:1984}
\begin{equation}
ds^2 =
-\frac{\Delta}{r^2}dt^2+\frac{r^2}{\Delta}dr^{2}+r^2(d\theta^2+\sin^2
\theta d\phi^2), 
\ \ \ \ \ \ \  \Delta = r^2-2Mr+Q^2= (r-r_+)(r-r_{-}), 
\end{equation}
and in the Coulomb gauge the only non-vanishing component of the
electromagnetic potential $A_{\mu}$ is $A_{0} = -\frac{Q}{r}$. The
roots of $\Delta$ will be called $r_+$ and $r_-$. For $Q<M$, we have
the non-extremal RN black hole and $r_+$ and $r_-$ are the event and
Cauchy horizons of the black hole, respectively, with $r_+>r_-$. When
$Q=M$, the black hole is extremal, where the roots are equal, $r_+ =
r_-$. In turn, for $Q>M$, the black hole space-time has a naked
singularity, with $\Delta$ having no real root. 

Linear perturbations of massless charged scalar ($s=0$) and spinor
($s=\frac{1}{2}$) fields in the RN background can be effectively
unified in a scheme similar to the radial Teukolsky master equation
(TME) for the Kerr black hole \cite{Teukolsky:1973ha}. The master
equation can be derived by separating the Klein-Gordon and the Dirac
equations in the RN background \cite{LivingRev.Rel.2:21999,slavyanov2000}, arriving at 
\begin{equation}
  \Delta^{-s}\frac{d}{dr}\bigg(\Delta^{1+s}\frac{dR_{s}(r)}{dr}
  \bigg)+\bigg(\frac{K(r)^{2}-2is(r-M)K(r)}{\Delta}+4is\omega
  r-2isqQ-{}_{s}\lambda_{\ell}\bigg)R_{s}(r)=0 
  \label{eq:radeq}
\end{equation}
where, $s=0,\pm 1/2$, $K(r) = \omega r^2-qQ r$ and
${}_{s}\lambda_{\ell}=(\ell-s)(\ell+s+1)$ is the separation constant. The
angular dependence of the solutions is given by the usual
spin-weighted spherical harmonics $Y^{s}_{\ell m}(\theta,\phi)$, where
the parameters $\ell$ and $m$ satisfy the usual constraints: $\ell$ is
a non-negative integer and a non-negative half-integer for bosonic and
fermionic case, respectively,  with $\ell \geq |s|$. For $m$, one has
essentially $-\ell \leq m \leq \ell$
\cite{doi:10.1063/1.1705135,Berti:2005gp}.

Quasinormal modes (QNMs) are in general an useful tool to
understand physical properties of black hole backgrounds, such as
linear stability. For the RN black hole, QNMs are solutions of
\eqref{eq:radeq} that correspond to purely outgoing waves at the outer
horizon and purely ingoing at infinity. The spectrum of these boundary
conditions is discrete, with complex frequencies, the real part
representing the oscillation character of the perturbations and the
imaginary part the damping factor.  The frequencies are not only 
functions of the mass $M$ and charge $Q$ of the black hole, but also
of the charge $q$ and spin $s$ of the perturbing field. The spectrum
is labelled by integers $m$, $\ell$ and $n$ with usual
interpretations.

The study of electromagnetic and gravitational QNMs in RN space-time
was considered in \cite{Kokkotas:1988fm}, using a semiclassical (WKB)
approach. In \cite{PhysRevD.41.2986}, Leaver considered the same problem
using the solution of the 3-term recurrence equation using continued
fraction (CF) method that he introduced in \cite{Leaver:1985ax}. The
CF method allowed for improved accuracy when compared with the WKB
approximation, but the numerics suffered for the near-extremal case
$Q\rightarrow M$. The continued fraction method was later improved by
\cite{Onozawa:1995vu}, and used to study the extremal case, obtaining
values for the massless scalar ($s=0$), electromagnetic and
gravitational cases when $Q=M$. 

For spinoral ($s=1/2$) perturbations, the initial analysis was made in
\cite{Wu:2004vb}, approximating the interaction between the field and
the black hole by a Pöschl-Teller potential. In
\cite{Richartz:2014jla}, the CF method was applied for the
non-extremal RN black hole to obtain the QNMs frequencies for spins $0$
and $1/2$, and various values of $qQ$. This analysis was extended for
extremal RN black holes in \cite{Richartz:2015saa}, using the same
strategy developed in \cite{Onozawa:1995vu}. In these analysis, the
study of the extremal limit, where $Q\rightarrow M$ is hindered by
numerical accuracy of the CF method, and the behavior the QNMs as one
approaches the extremal point is still in need of clarification.

In a series of papers \cite{daCunha:2015ana,CarneirodaCunha:2019tia},
building on previous work
\cite{Motl:2003cd,Neitzke:2003mz,Castro2013b}, properties
of the scattering of the fields were related to the monodromy
properties of the solutions of the confluent Heun differential
equation governing the black hole perturbations. In turn, these
monodromy properties are obtainable from the parameters of the
differential equation by means of the Riemann-Hilbert map, and more
conveniently expressed in terms of the Fifth Painlevé transcendent
tau function $\tau_V$. By using the general expansion of $\tau_V$ 
given in \cite{Gamayun:2013auu,Lisovyy:2018mnj}, one achieves an
analytical solution to the scattering problem, given implicitly in
terms of transcendental equations which can be studied
numerically. The relation between these expansions and the properties 
of conformal blocks both in the $c=1$ \cite{CarneirodaCunha:2019tia}
and semiclassical cases \cite{Bershtein:2021uts} hints at a possible
underlying connection between black hole physics and conformal field
theories. At any rate, the procedural solution of the monodromy
problem in terms of conformal blocks provide us with a numerical
recipe that is not encumbered by the drawbacks of the CF method. In
\cite{daCunha:2021jkm} the $\tau_V$ method was used to study
the QNMs frequencies for the Kerr black hole, focussing particularly
in the extremal limit.

Our main objective in this work is to apply the isomonodromic
method to study the QNMs of scalar and spinorial massless
perturbations of the RN black hole. We will treat the black hole and
field charges, respectively $Q$ and $q$, generically, but will focus
on the extremal limit. The plan of the paper is as follows: in Section
\ref{sec:monos} we revise the RH method, and in Section
\ref{sec:Qgeneric} we derive the relevant parameters for the confluent
Heun equation associated to \eqref{eq:radeq} and present the results
of the numerical analysis for $Q<M$ and $Q\rightarrow M$
separately. In Section \ref{sec:Conflimit_and_RNextremal} we deal with
the extremal case separately. We close by commenting the results in
Section \ref{sec:discussion} and presenting some future prospects. For
completeness, we included an Appendix \ref{sec:tools} with the
relevant technical details on the tau functions and monodromy
properties of the solutions.

\section{Monodromy Properties}
\label{sec:monos}

Let us revise the isomonodromic method used in
\cite{CarneirodaCunha:2019tia,daCunha:2021jkm}. The radial
equation \eqref{eq:radeq} is a particular case of the confluent Heun
differential equation
\begin{equation}
  \frac{d^2
   y}{dz^2}+\bigg[\frac{1-\theta_0}{z}+\frac{1-\theta_{t_0}}{z-t_0}
   \bigg]\frac{dy}{dz}-\bigg[\frac{1}{4}
     +\frac{\theta_{\star}}{2z}+\frac{t_0c_{t_0}}{z(z-t_0)}\bigg]y(z)=0, 
  \label{heuneq}
\end{equation}
defined for complex $z$, with regular singular behavior at $0$ and
$t_0$ and an irregular singularity of Poincaré rank 1 at $z=\infty$.

The Riemann-Hilbert map is defined from the accessory parameter
$c_{t_0}$ and conformal modulus $t_0$ of \eqref{heuneq} to the
\textit{monodromy data} $\{\sigma, \eta \}$ of its solutions. Let us briefly
review the isomonodromy method to compute this monodromy data as
described in \cite{CarneirodaCunha:2019tia}. This consists of
rephrasing the confluent Heun equation as a matrix system
\begin{equation}
  \frac{d}{dz}\Phi(z) = \mathbf{A}(z)\Phi(z),\qquad
  \Phi(z) = \begin{pmatrix}
    y_+(z) & y_-(z) \\
    w_+(z) & w_-(z)
  \end{pmatrix},
  \label{eq:matrixsys}
\end{equation}
where $y_\pm(z)$ are two linearly independent solutions of
\eqref{heuneq} and $w_\pm(z)$ two auxiliary functions, obtained from
$y_\pm(z)$ and derivatives multiplied by rational functions. The ``gauge
potential'' $\mathbf{A}(z)$ has a partial fraction expansion
\begin{equation}
  \mathbf{A}(z)=\frac{1}{2}\sigma_3 + \frac{\mathbf{A}_0}{z}+
  \frac{\mathbf{A}_{t}}{z-t},
  \label{eq:gaugepot}
\end{equation}
where $\sigma_3=\left(
  \begin{smallmatrix} 1 & 0 \\ 0 & -1 \end{smallmatrix}\right)$ is the
Pauli matrix and the coefficients $\mathbf{A}_0$ and
$\mathbf{A}_{t_0}$ are independent of $z$. In this expansion, we are
allowing for the existence of gauge transformations which alter the
value of the conformal modulus $t_0$.

Monodromy properties of the solutions $y_\pm(z)$, or equivalently of
$\Phi(z)$, are in this way translated intro holonomy properties of
$\mathbf{A}(z)$. For instance, the monodromy parameter $\sigma$ is
related to a non-contractible path $\gamma$ circling singularities at
\eqref{heuneq} at both $z=0$ and $z=t$. Thus
\begin{equation}
  \Phi(z_\gamma)=\Phi(z)\mathbf{M}_{\gamma},\qquad
  \mathbf{M}_{\gamma}=\mathrm{P\exp}\left[
    \oint_{\gamma}\mathbf{A}(z)dz \right],
\end{equation}
where $\mathrm{P\exp}$ stands for path-ordered exponentiation, and
$z_\gamma = (z-z_0)+z_0e^{2\pi i}$, where $|z_0|>|t|$. The
gauge-invariant quantity $\sigma$ is defined as the trace of 
the monodromy matrix $2\cos\pi\sigma = \Tr
\widehat{\mathbf{M}}_{\gamma}$, where $\widehat{\mathbf{M}}_{\gamma}$
is defined from $\mathbf{M}_{\gamma}$ by factoring out the abelian
$\mathrm{U(1)}$ diagonal factor. The monodromy parameter $\eta$ has a
more involved definition in terms of the potential $\mathbf{A}(z)$,
and we refer to \cite{CarneirodaCunha:2019tia} for details.

The calculation of $\sigma,\eta$ from the accessory parameter
$c_{t_0}$ and the conformal modulus $t_0$ depends on a way of
embedding the differential equation \eqref{heuneq} into the matrix
system \eqref{eq:matrixsys}, defining a gauge potential
$\mathbf{A}(z)$ (almost) uniquely. The Riemann-Hilbert map is made
possible by the \textit{isomonodromic $\tau$ function},  defined by
the groundbreaking papers of the Kyoto school
\cite{Jimbo:1981aa,Jimbo:1981ab,Jimbo:1981ac}, 
\begin{equation}
  \frac{\partial}{\partial t}\log \tau_V = \frac{1}{2}\Tr
  \sigma_3\mathbf{A}_{t} +\frac{1}{t}\Tr \mathbf{A}_0
  \mathbf{A}_{t},
  \label{eq:taufunction}
\end{equation}
which has its existence guaranteed by the invariance of monodromy data
like $\sigma,\eta$ under gauge transformations of $\mathbf{A}(z)$
keeping the partial fraction expansion \eqref{eq:gaugepot}. The most
relevant factor for stating the map is the discovery that $\tau_V$ has
a natural expansion in terms of monodromy data \cite{Jimbo:1982aa}.

In terms of $\tau_V$, the Riemann-Hilbert map between $t_0,c_{t_0}$
and $\sigma,\eta$  is implicitly expressed by the following equations 
\begin{equation}
  \tau_V(\vec{\theta};\sigma,\eta;t_0)=0, \qquad
  t_0\frac{d}{dt}\log \tau_V(\vec{\theta}_{-};\sigma-1,\eta;t_0)
  -\frac{\theta_0(\theta_t-1)}{2} = t_0 c_{t_0},
  \label{RHmap}
\end{equation}
where $\vec{\theta}=\{\theta_0,\theta_{t},\theta_{\star}\}$ are the
parameters in \eqref{heuneq} associated to the local monodromy of
solutions.

The first equation is the Painlevé V version of the Toda
equation, studied in \cite{Okamoto:1987aa}, and was shown in
\cite{CarneirodaCunha:2019tia} to be related to the well-posedness of
the initial value problem for the Painlevé V transcendent
equation. The monodromy arguments of the second equation are shifted:
$\vec{\theta}_{-} =\{\theta_0, \theta_{t}-1,\theta_{\star}+1\}$, and
defines the accessory parameter $c_{t_0}$ as the logarithm derivative
of the isomonodromic tau function. We refer to \cite{CarneirodaCunha:2019tia} for
details on the relation between $\tau_V$, the fifth Painlevé
transcendent and the theory of isomonodromic deformations. 

The role of the so-called composite monodromy parameters $\{
\sigma,\eta\}$ was further studied in \cite{Lisovyy:2018mnj}, in the context
of confluent conformal blocks. In \cite{Lisovyy:2021bkm} (see also
\cite{daCunha:2021jkm}), it was noted that $\sigma$ parametrizes the
so-called Floquet solutions of \eqref{heuneq}, and a direct relation
between $\sigma$ and $c_{t_0}$ was derived, 
\begin{multline}
t_0c_{t_0}=\frac{(\sigma-1)^2-(\theta_t+\theta_0-1)^2}{4}+
\frac{\theta_\star(\sigma(\sigma-2)+\theta_t^2-\theta_0^2)}{
	4\sigma(\sigma-2)}t_0 \ +\\
+\left[\frac{1}{32}+\frac{\theta_\star^2(\theta_t^2-\theta_{0}^2)^2}{64}
\left(\frac{1}{\sigma^3}-\frac{1}{(\sigma-2)^3}\right)
+\frac{(1-\theta_\star^2)(\theta_0^2-\theta_{t}^2)^2+2\theta_\star^2
	(\theta_0^2+\theta_{t}^2)}{32\sigma(\sigma-2)}\right.
\\ \left.-
\frac{(1-\theta_\star^2)((\theta_0-1)^2-\theta_{t}^2)((\theta_0+1)^2-
	\theta_{t}^2)}{32(\sigma+1)(\sigma-3)}\right]t_0^2+
{\cal O}(t_0^3).
\label{eq:c5expansion}
\end{multline}
This formula can be obtained by various different methods, and the
equivalence between the semiclassical confluent conformal blocks
approach of \cite{Litvinov:2013sxa,Lisovyy:2021bkm,Bershtein:2021uts}
and the $c=1$ isomonodromic approach of
\cite{Lisovyy:2018mnj,CarneirodaCunha:2019tia} is a non-trivial
consequence of the exponentiation property of conformal blocks.

The practical use of the Riemann-Hilbert map \eqref{RHmap} relies on
effective ways to compute $\tau_V$. In a seminal work
\cite{Jimbo:1982aa}, Jimbo showed that the isomonodromic $\tau$
functions can be conveniently expanded in terms of the monodromy
parameters $\{\sigma,\eta\}$, and, in \cite{Gamayun:2013auu}, the
authors used the relation between isomonodromic deformations,
conformal blocks and Seiberg-Witten theory to propose the full
expansion of $\tau_V$ in terms of Nekrasov functions. Later, in
\cite{Lisovyy:2018mnj}, the Painlevé V $\tau$ function was recast in
terms of the Fredholm determinant associated to the actual
Riemann-Hilbert problem of finding analytic functions on the complex
plane with prescribed jumps. The latter formulation of $\tau_V$ is
particularly suited for numerical studies, as the authors have
conducted for the fundamental QNMs for generic spin perturbations of
the Kerr black holes in \cite{daCunha:2021jkm}, where the reader can
find complete formulas. For our purposes, the small-$t$ expansion of
the $\tau_{V}$ will suffice \eqref{eq:fredholmV}. For $\sigma$ in the
fundamental domain $-1<\Re(\sigma)<1$, we have 
\begin{multline}
  \tau_V(\vec{\theta};\sigma,\eta;t)=
  C_{V}(\vec{\theta};\sigma)
  t^{\frac{1}{4}(\sigma^2-\theta_0^2-\theta_t^2)}
  e^{\frac{1}{2}\theta_t t}\bigg(
1-\left(\frac{\theta_t}{2}-\frac{\theta_\star}{4}
+\frac{\theta_\star(\theta_0^2-\theta_t^2)}{4\sigma^2}\right)t
\\ -
\frac{(\theta_\star+\sigma)((\sigma+\theta_t)^2-
	\theta_0^2)}{8\sigma^2(\sigma-1)^2}\kappa_V^{-1}
t^{1-\sigma}-
\frac{(\theta_\star-\sigma)((\sigma-\theta_t)^2-
	\theta_0^2)}{8\sigma^2(\sigma+1)^2}
\kappa_V\, t^{1+\sigma}+{\mathcal O}(t^2,|t|^{2\pm
	2\Re\sigma})\bigg),
\label{eq:expansiontauV}
\end{multline}
where $C_V$ is an arbitrary (non-zero) constant and
\begin{equation}
\kappa_V =e^{i\pi\eta}\Pi_{V}=
e^{i\pi\eta}\frac{\Gamma(1-\sigma)^2}{\Gamma(1+\sigma)^2}
\frac{\Gamma(1+\tfrac{1}{2}(\theta_\star+\sigma))}{
	\Gamma(1+\tfrac{1}{2}(\theta_\star-\sigma))}
\frac{\Gamma(1+\tfrac{1}{2}(\theta_t+\theta_0+\sigma))
	\Gamma(1+\tfrac{1}{2}(\theta_t-\theta_0+\sigma))}{
	\Gamma(1+\tfrac{1}{2}(\theta_t+\theta_0-\sigma))
	\Gamma(1+\tfrac{1}{2}(\theta_t-\theta_0-\sigma))}.
\label{eq:kappaV}
\end{equation}

In general, scattering amplitudes can be written in terms of the
monodromy parameters \cite{daCunha:2015ana,CarneirodaCunha:2019tia},
so by solving the Riemann-Hilbert map one can solve the scattering
problem. For the calculation of quasinormal modes, we must impose
boundary conditions such that there is no energy flux out of the black
hole event horizon at $r_+$ and no energy flux out at infinity. In
\cite{daCunha:2021jkm}, the authors showed that these boundary
conditions can be cast in terms of monodromy data as
\begin{equation}
e^{i\pi\eta}=e^{-i\pi\sigma}
\frac{\sin\tfrac{\pi}{2}(\theta_\star+\sigma)}{
	\sin\tfrac{\pi}{2}(\theta_\star-\sigma)}
\frac{\sin\tfrac{\pi}{2}(\theta_t+\theta_0+\sigma)
	\sin\tfrac{\pi}{2}(\theta_t-\theta_0+\sigma)}{
	\sin\tfrac{\pi}{2}(\theta_t+\theta_0-\sigma)
	\sin\tfrac{\pi}{2}(\theta_t-\theta_0-\sigma)}.
\label{eq:quantizationV}
\end{equation}
This condition is equivalent to imposing that the connection matrix
$C_t$ between local solutions of \eqref{heuneq} constructed around the
$z=t_0$ and local solutions at $z=\infty$ is lower triangular, see
part \ref{sec:quantcond} of the Appendix. 

With the explicit formulation of $\tau_V$, the Riemann-Hilbert map
\eqref{RHmap} can in principle be used to compute the monodromy
parameters $\{\sigma,\eta\}$ given the parameters of the confluent
Heun equation \eqref{heuneq}. Given generic parameters, the QNM
condition \eqref{eq:quantizationV} overdetermines the system and thus
will allow for solutions only for a discrete set of values of $t_0$
and $c_{t_0}$. Operationally speaking, one can use the expansion of
$\tau_V$ \eqref{eq:expansiontauV} to solve the first equation in
\eqref{RHmap} to compute $\eta$ in terms of $\sigma$, $t_0$ and the
$\vec{\theta}$, and the expansion of the accessory parameter
\eqref{eq:c5expansion} to compute $\sigma$ in terms of the parameters
of \eqref{heuneq}. 

Following this strategy, one notes from \eqref{eq:expansiontauV} that
$\tau_V$ is meromorphic in $\kappa_V t^{\sigma}$ so the equation
$\tau_{V}(\vec{\theta},\sigma,\eta,t_0)=0$ can be inverted to define a
series for $\kappa_V t^{\sigma}$, or $e^{i\pi\eta}$ in terms of
$t_0$. Assuming that $\sigma$ satisfies $0<\Re(\sigma)<1$, we 
arrive at the expression 
\begin{equation}
  \Theta_{V}(\vec{\theta};\sigma)e^{i\pi\eta}t_{0}^{\sigma_0-1} =
  \chi_{V}(\vec{\theta};\sigma;t_0),
  \label{eq:zerotau5p}
\end{equation} 
with $\Theta_{V}(\vec{\theta};\sigma)$ expressed in terms of ratios of
gamma functions 
\begin{equation}
\Theta_V(\vec{\theta};\tilde{\sigma})=
\frac{\Gamma^2(2-\sigma)}{\Gamma^2(\sigma)}
\frac{\Gamma(\tfrac{1}{2}(\theta_\star+\sigma))}{
	\Gamma(1+\tfrac{1}{2}(\theta_\star-\sigma))}
\frac{\Gamma(\tfrac{1}{2}(\theta_t+\theta_0+\sigma))}{
	\Gamma(1+\tfrac{1}{2}(\theta_t+\theta_0-\sigma))}
\frac{\Gamma(\tfrac{1}{2}(\theta_t-\theta_0+\sigma))}{
	\Gamma(1+\tfrac{1}{2}(\theta_t-\theta_0-\sigma))}
\label{eq:theta5}
\end{equation}
and the function $\chi_{V}(\vec{\theta};\sigma;t_0)$ analytic for
$t_0$ small, with expansion 
\begin{multline}
\chi_V(\vec{\theta};\tilde{\sigma};t_0)
=1+(\tilde{\sigma}-1)\frac{\theta_\star
	(\theta_t^2-\theta_0^2)}{\tilde{\sigma}^2(\tilde{\sigma}-2)^2}t_0+
\left[\frac{\theta_\star^2(\theta_t^2-\theta_{0}^2)^2}{64}
\left(\frac{5}{\tilde{\sigma}^4}-\frac{1}{(\tilde{\sigma}-2)^4}
-\frac{2}{(\tilde{\sigma}-2)^2}+\frac{2}{\tilde{\sigma}(\tilde{\sigma}-2)}\right)
\right.\\
-\frac{(\theta_t^2-\theta_{0}^2)^2+2\theta_\star^2
	(\theta_t^2+\theta_{0}^2)}{64}\left(
\frac{1}{\tilde{\sigma}^2}-\frac{1}{(\tilde{\sigma}-2)^2}\right)
\\ \left.
+\frac{(1-\theta_\star^2)(\theta_t^2-(\theta_{0}-1)^2)(\theta_t^2
	-(\theta_{0}+1)^2)}{128}\left(\frac{1}{(\tilde{\sigma}+1)^2}-
\frac{1}{(\tilde{\sigma}-3)^2}\right)\right]t_0^2+{\cal O}(t_0^3).
\label{eq:chi5}
\end{multline}
For $\Re(\sigma)<0$, we also have an expression similar for
\eqref{eq:zerotau5p}, where one simply changes to $\sigma \rightarrow
-\sigma$ and $e^{i\pi\eta} \rightarrow e^{-i\pi\eta}$. Although it
will not be necessary for our purposes, one can in principle compute
the expansion of $\chi_V$ to arbitrary order in $t_0$.

Finally, we can incorporate the condition \eqref{eq:quantizationV}
into the expression \eqref{eq:zerotau5p}. Using the definition of
$\Theta_V$, and the identity $\Gamma(z)\Gamma(1-z) =\pi/\sin(\pi z)$,
one obtains, when $\eta$ satisfies \eqref{eq:quantizationV}, 
\begin{equation}
\Theta_V(\vec{\theta},\sigma) e^{i\pi\eta} =
-e^{-i\pi\sigma}\Theta_V(-\vec{\theta},\sigma),
\end{equation}
Substituting back into
\eqref{eq:zerotau5p}, we find
\begin{equation}
  -e^{-i\pi\sigma}\Theta_V(-\vec{\theta},\sigma)
  t_0^{\sigma-1}=\chi_V(\vec{\theta};\sigma;t_0),
  \label{eq:zerochi5}
\end{equation}
again assuming $\Re(\sigma)>0$. For $\Re(\sigma)<0$, one finds the
right expansion by simply sending $\sigma\rightarrow -\sigma$ in
\eqref{eq:zerochi5}. With the expression \eqref{eq:zerochi5}, one can
then consider for the QNM problem the overdetermined system of
equations consisted by \eqref{eq:c5expansion} and \eqref{eq:zerochi5},
with $\sigma$ the free parameter.

\section{Radial System and Numerical Results for $0\leq Q \leq M$}
\label{sec:Qgeneric}

After these generic comments, let us consider the RH map
\eqref{RHmap} for the radial equation \eqref{eq:radeq}. The radial
equation \eqref{eq:radeq} can be brought to the standard form
\eqref{heuneq} by the change of variables 
\begin{equation}
	R(r)=(r-r_{-})^{-(s+\theta_{-})/2}(r-r_{+})^{-(s+\theta_{+})/2}y(z),
        \qquad z=2i\omega(r-r_{-})
\end{equation}
where	
\begin{equation}
  \begin{gathered}
    \theta_{-}=  s+\frac{i}{2\pi
      T_{-}}\bigg(\omega-\frac{qQ}{r_{-}}\bigg), \qquad \theta_{+}=
    s+ \frac{i}{2\pi
      T_{+}}\bigg(\omega-\frac{qQ}{r_{+}}\bigg), \qquad
    \theta_{\star}=-2s+2i(2M\omega -qQ),\\
    2\pi T_{\pm} = \frac{r_{\pm}-r_{\mp}}{2 r^{2}_{\pm}},   \qquad
    r_{\pm} = M\pm\sqrt{M^2 -Q^2}.
    \label{parameters}
\end{gathered}
\end{equation}
This simple transformation can also be found in
\cite{Fiziev:2009wn}. From it, one can find the accessory parameter
and the modulus 
\begin{equation}
    z_0c_{z_0} =   {}_{s}\lambda_{l,m}
    +2s-i(1-2s)qQ+(2qQ+i(1-3s))\omega r_++i(1-s)\omega r_-
    -2\omega^2 r_+^2, \qquad
    z_0= 2i\omega(r_+ - r_-).
    \label{modacce}
\end{equation}
In the following it will be convenient to parametrize $Q/M = \cos\nu$
with $\nu \in [0,\pi/2]$, with the extremal limit $r_-\rightarrow r_+$
corresponding to $\nu\rightarrow 0$. In terms of $\nu$ , the monodromy
parameters \eqref{parameters} are rewritten as
\begin{equation}
  \begin{gathered}
    \theta_{-} = -s -\frac{i}{2\pi
      T_{-}}\bigg(\omega-\frac{qQ}{M(1-\sin\nu)}\bigg),
    \qquad \theta_{+} = s + \frac{i}{2\pi
      T_{+}}\bigg(\omega-\frac{qQ}{M(1+\sin\nu)}\bigg), \qquad
    \theta_{\star} = -2s+2i(2M\omega-qQ)
    \label{eq:nuparametrization} \\ 2\pi T_{\pm}
    =\pm\frac{\sin\nu}{M(1\pm\sin\nu)^2}, \qquad
    r_\pm =M(1\pm \sin\nu),
  \end{gathered}
\end{equation}
and the accessory parameters\eqref{modacce}
\begin{equation}
  \begin{gathered}
    z_0c_{z_0}= {}_{s}\lambda_{l,m} +2s-i(1-2s)qQ +
    2(qQ+i(1-2s)+(qQ-is)\sin\nu) M\omega
    -2(1+\sin\nu)^2(M\omega)^2, \\
    z_0= 4iM\omega\sin\nu .
  \end{gathered}
  \label{modaccASnufunction}
\end{equation}

For generic $Q$, the QNM calculation follows the
procedure defined in \eqref{RHmap}, written in terms of
the parameters \eqref{parameters} and \eqref{modacce} as 
\begin{equation}
\tau_V(\vec{\theta}_{\text{non-ext}};\sigma,\eta;z_{0})=0,
\qquad
z_{0}\frac{d}{dt}\log\tau_V(\vec{\theta}_{\text{non-ext},-};
\sigma-1,\eta;z_{0})-
\frac{\theta_{-}(\theta_{+}-1)}{2}=
z_{0}c_{z_0},
\label{eq:radialsystemeqn}
\end{equation}
with
$\vec{\theta}_{\text{non-ext}}=\{\theta_{-},\theta_{+},\theta_{\star}
\}$,
$\vec{\theta}_{\text{non-ext}}=\{\theta_{-},\theta_{+}-1,\theta_{\star}+1
\}$, the modulus and the accessory parameter given by
\eqref{eq:c5expansion}. To distinguish from the extremal parameters we
will introduce later, we have included the subscript
$\text{non-ext}$. For the actual numerical analysis, it is more
convenient to work directly with the system
\eqref{eq:radialsystemeqn}, rather than using the analysis at the end
of Section \eqref{sec:monos}.

\subsection{Numerical Results for $\ell >|s|$}

We are now ready to employ the technique described in the previous
sections to determine the fundamental modes for spin-$0$ and
spin-$\frac{1}{2}$ perturbations of the non-extremal RN black hole. 
The initial analysis is concentrated on the case $\ell > |s|$, which
means $\ell=1,2,3,\ldots$ for the scalar case and
$\ell=3/2,5/2,\ldots$ for the spinorial case.  As explained in
\cite{Chang2007MassiveCQ}, the potential terms for spin-$1/2$ produce
the same QNMs spectrum for both values $s = \pm1/2$ and there is no loss
of generality in considering only the $s = -1/2$ case. Additionally,
since the RN black hole is symmetric with respect to transformations
$q\rightarrow -q$ and $\omega \rightarrow -\omega^{*}$, we consider
only positive $q$ and, consequently, $\Re(M\omega)>0$.
\begin{figure}[htb]
  \begin{center}
    \includegraphics[width=0.95\textwidth]{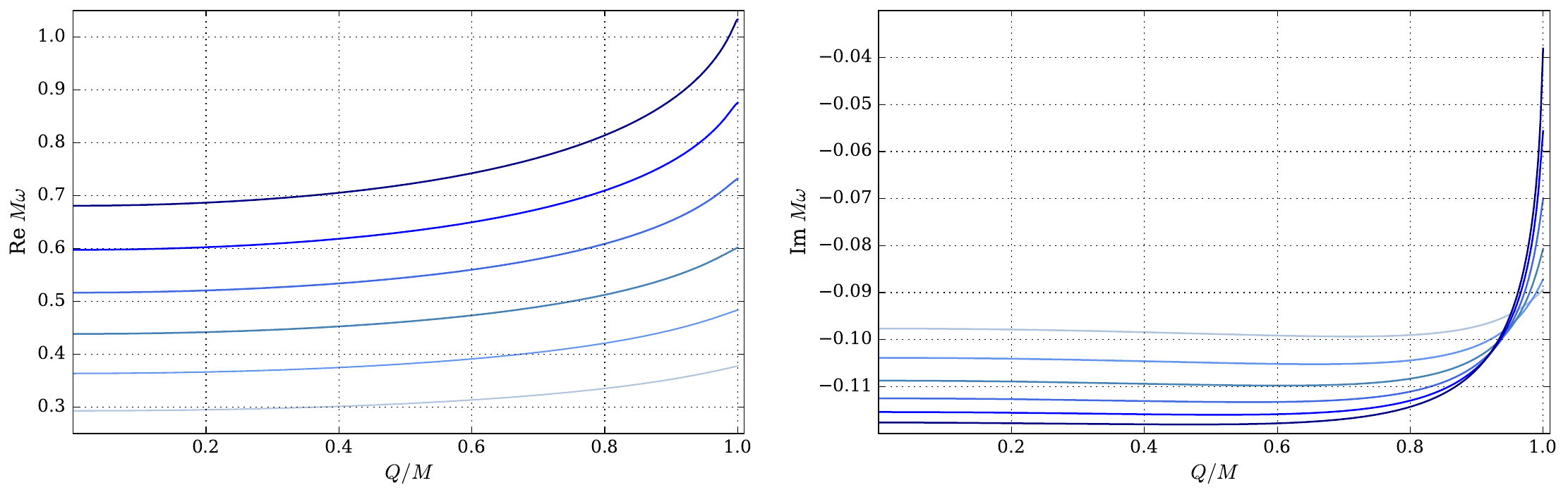}
    \includegraphics[width=0.95\textwidth]{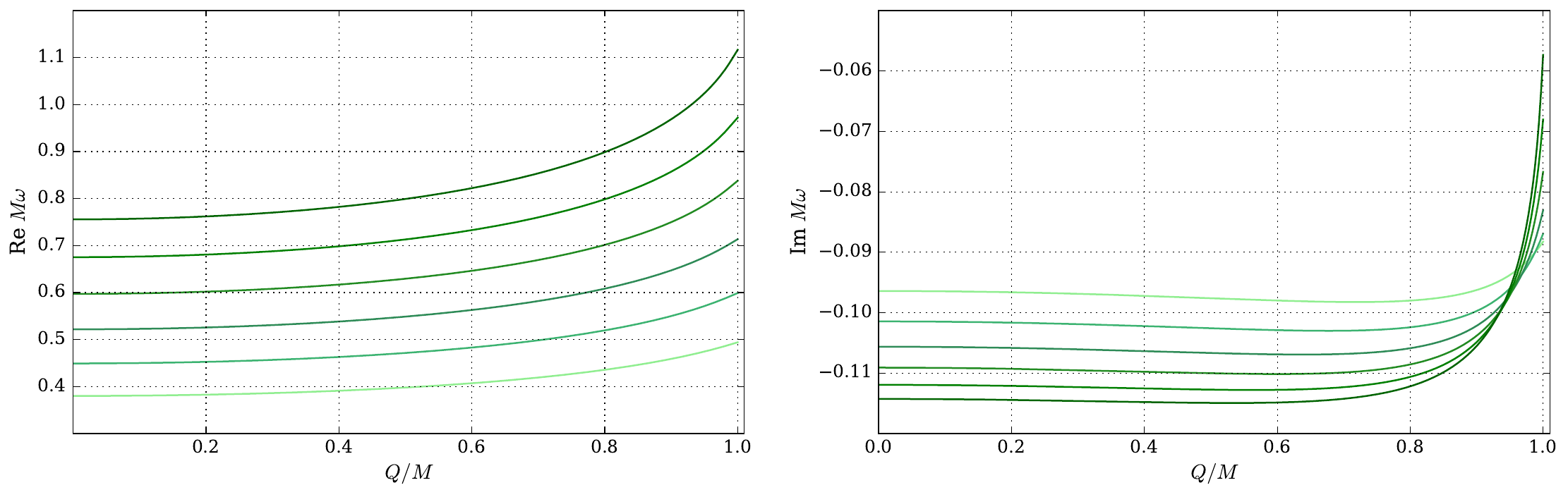}
  \end{center}
  \caption{Fundamental modes calculated for scalar and spinor
    perturbations in non-extremal RN black hole. Results for  $s=0,
    l=1$ (top) and $s=-\frac{1}{2}, \ell=\frac{3}{2}$ (bottom) as a
    function of $Q/M$ for $qQ= 0.0$ (lightest), $0.2,0.4,0.6,0.8$ and
    $1.0$ (darkest).} 
  \label{fig:s0l1}
\end{figure}

In Fig. \eqref{fig:s0l1}, we show the QNMs frequencies for $s=0$,
$\ell=1$ and $s=-\frac{1}{2}$, $\ell= \frac{3}{2}$ as a function of
$qQ$. These values were obtained by directly solving
\eqref{eq:radialsystemeqn}, using a numeric implementation of
\eqref{eq:expansiontauV} and \eqref{eq:c5expansion} in
\href{http://julialang.org}{Julia} language, finding the roots of
$\tau_V$ using a simple root-finding algorithm (Muller's method). The
Fredholm determinant involved in the definition of $\tau_V$ was
truncated at $N_f=32$ Fourier components and \eqref{eq:c5expansion}
at $N_c=64$ convergents. The method converges fastly due to Miwa's
theorem \cite{Miwa:1981aa}, which asserts that $\tau_V$ only has
isolated and simple zeros away from its critical values $t=0,\infty$. However,
the zero locus structure can become very intricate, particularly close to
extremality, so when varying $\frac{Q}{M}$ we use the QNM value for a
particular point as the guess for the following point. The
implementation of the Painlevé transcendents $\tau_{VI}$, $\tau_{V}$
and $\tau_{III}$ based on the Fredholm determinants is available at
\cite{Github}.

When solving the set of equations \eqref{eq:radialsystemeqn}, we have
verified that the fundamental modes for $qQ=0$, $Q/M =0$, in the cases
$s=0$, $l=1,2,3,...$, and $s=-\frac{1}{2}$,
$l=\frac{3}{2},\frac{5}{2},\frac{7}{2},...$, are in agreement with the
modes for Schwarzschild black hole calculated using the WKB
approximation \cite{Konoplya:2004ip,Cho:2003qe} and CF method in
the \href{https://pages.jh.edu/eberti2/ringdown/}{literature}, see Table
\ref{fig:s0l1}. In turn,  in the case of the
non-extremal RN black hole with $qQ\neq 0$ and $Q/M=0.999$, we
recovered the fundamental modes for $s=0$, $l=1,2,3$ and
$s=-\frac{1}{2}$, $\ell=\frac{3}{2},\frac{5}{2},\frac{7}{2}$
calculated using the modified version of the continued fraction method
and listed in \cite{Richartz:2015saa}. Given that the CF method has
convergence problems near the extremality, we have extended the
analysis by  solving the system \eqref{eq:radialsystemeqn} for
$Q\rightarrow M$, as we can seen in Fig. \ref{fig:s0l1}. By employing
the procedure described above, we can consider values up to $Q/M =
1-10^{-11}$ in reasonable time and accuracy. One notes that the method
infers a smooth extremal limit for these modes, a point we will return to
below. 

\begin{table}[htb]
  \begin{tabular}{|c|c|c|c|c|}
    \hline $s$& $\ell$ & 
                         ${}_{s}\omega_{\ell}$\, ($\tau_V
                         \text{ function}$)
                         &  ${}_{s}\omega_{\ell}$\,
                           ($\text{CF method}$)  \\
    \hline
    0 &1  & $0.292936133267 - 0.097659988914i$   &   $ 0.292936133267
                                                   - 0.097659988913i $
    \\ \hline
    0 &2  & $0.483643872211 - 0.096758775978i$   &  $ 0.483643872211 -
                                                   0.096758775978i $
    \\ \hline
    0 &3  & $0.675366232537 - 0.096499627734i$   &  $ 0.675366232537 -
                                                   0.096499627734i $
    \\ \hline
    $-\frac{1}{2}$ & $\frac{3}{2}$ &$0.380036764833 - 0.096405208085i$
                         &  $0.380036764833 - 0.096405208085i $  \\
    \hline
    $-\frac{1}{2}$ & $\frac{5}{2}$  & $0.574093974298 -
                                      0.096304784939i$  &
                                                          $0.574093974298 - 0.096304784939i$   \\ \hline
    $-\frac{1}{2}$ & $\frac{7}{2}$  & $ 0.767354592773 -
                                      0.096269878994i$  &
                                                          $0.767354592773 - 0.096269878994i$   \\ \hline
  \end{tabular}
  \caption{To the left, the fundamental modes for Schwarzschild black
    hole $q=Q=0$ recovered with $\ell>|s|$. For comparison, we show to
    the right the values in the
    \href{https://pages.jh.edu/eberti2/ringdown/}{literature}
    \cite{Berti:2005gp}.} 
  \label{tab:schwarzschild_s01over2}
\end{table}

\subsection{Numerical Results for $\ell =|s|$}

We now focus our attention in the case $\ell =|s|$ and investigate the
behavior of the fundamental modes as $Q/M$ and $qQ$ vary. In our
analysis, we have obtained QNMs for $Q$ going from $0$ to $M$ as a
function of $qQ$ and observed two different types of behavior in the
limite $Q\rightarrow M$, as follows:
\begin{enumerate}[I.]
\item
  For $qQ$ below a spin-dependent critical value $qQ_c(s)$ modes
  calculated from \eqref{eq:radialsystemeqn} behave similarly to $\ell
  > |s|$, with finite limits for the real and imaginary parts of the
  QNMs frequencies as  $Q\rightarrow M$. Following
  \cite{Richartz:2014jla}, we will call these ``damping modes''. 
\item
  For $qQ$ above the spin-dependent critical value $qQ_c(s)$, the
  frequency converges to $q$ as $Q\rightarrow M$ ($\nu\rightarrow
  0$). In this case, the imaginary part of the QNM frequency tends to
  zero, in a non-damping behavior. Also, the existence of a
  natural small parameter allows for the analytical treatment of these
  modes.
  \label{item:case}
\end{enumerate}

\begin{figure}[htb]
  \begin{center}
    \includegraphics[width=0.95\textwidth]{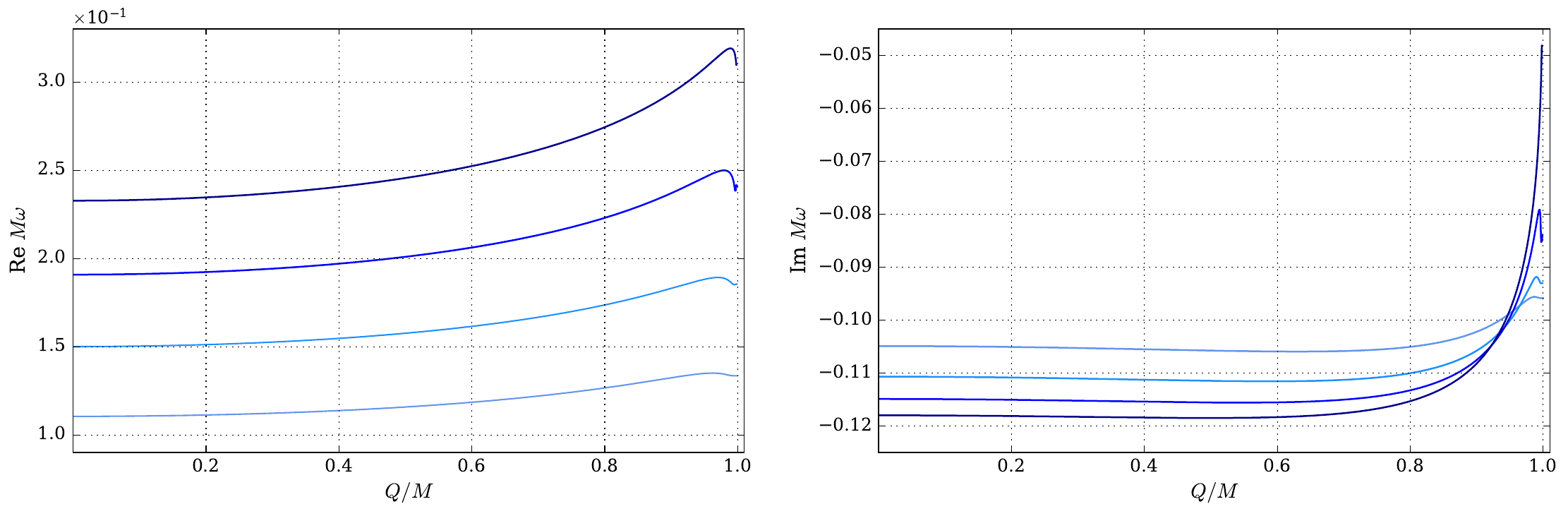}
    \includegraphics[width=0.95\textwidth]{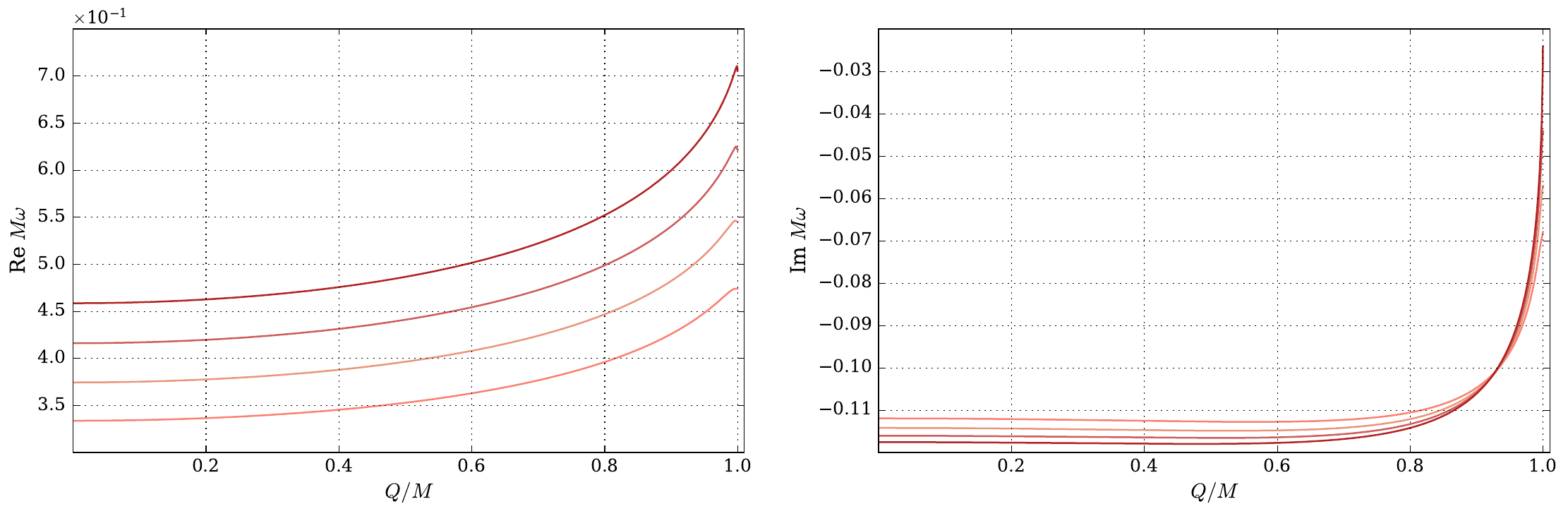}
  \end{center}
  \caption{Fundamental modes for  $s=0, l=0$ (top) and
    $s=\frac{1}{2},\ell = \frac{1}{2}$ (bottom) with $qQ$ varying from
    $0$ (lightest) to $0.3$ (darkest) with $0.1$ of increment.}
  \label{fig:s01over2l01over2}
\end{figure}

Before checking each case separately, let us illustrate in
Fig. \ref{fig:s01over2l01over2} the qualitative difference between
cases I and II in the usual graphs where $M\omega$ is plotted against
$Q/M$. For $s=\ell=0$, we note the appearance of non-trivial behavior
for $Q\rightarrow M$ as one increases $qQ$. In this regime, the
parametrization using $\nu$ \eqref{eq:nuparametrization} is more
adequate, so we will switch to it. Also, one can appreciate the
difficulty in studying these different behaviors using the usual CF
method, given that it does not converge easily (if at all) as one
approaches the extremal limit $\nu\rightarrow 0$.

The numerical analysis for these modes use the same procedure
described above: we simply solve \eqref{eq:radialsystemeqn} for
$\ell=|s|$. As a way of checking the procedure, we have recovered the
fundamental mode for $qQ=0$, $s=\ell=0$, and $Q/M=0$ obtained for
scalar perturbation in the Schwarzschild black hole and listed in the
\href{https://pages.jh.edu/eberti2/ringdown/}{literature}. Also the
fundamental mode for Dirac perturbation in this background was also
retrieved and compared with our implementation of the CF method. This
frequency was obtained in \cite{Konoplya:2004ip}, using the WKB
approximation which does not permit accurate comparisons. These values
are presented in Table \ref{tab:schwarzschilds=ell}. We have also
checked our results for $qQ\neq 0$ and $Q/M\lesssim 0.999$ against
\cite{Richartz:2015saa} and found excellent agreement.

As one can observe from Fig. \ref{fig:s01over2l01over2}, the extremal
limit $\nu\rightarrow 0$ of these modes shows non-trivial structure
depending on the value of the coupling $qQ$. Let us now consider this
limit in more detail. 

\begin{table}[htb]
  \begin{tabular}{|c|c|c|c|c|}
    \hline
    \ $s$& $\ell$ & ${}_{s}\omega_{\ell}$\,
                    ($\tau_V\text{ function}$) &
                                                ${}_{s}\omega_{\ell}$\, 
                                               ($\text{CF  method}$)  \\
    \hline
    \ $0$ & $0$ &  $0.110454939080 - 0.104895717087i$   &  $
                                                          0.110454939080
                                                          -
                                                          0.104895717087i
                                                          $  \\
    \hline
    $-\frac{1}{2}$ & $\frac{1}{2}$  & $0.182962870255 -
                                      0.096982392762i$  & $
                                                          0.182962870255
                                                          -
                                                          0.096982392762i
                                                          $   \\
    \hline
  \end{tabular}
  \caption{Comparison of the fundamental QNMs frequencies for
    Schwarzschild $q=Q=0$ black hole in the case $\ell=|s|$.}
  \label{tab:schwarzschilds=ell}
\end{table}

\subsection*{Non-damping modes for $\nu \rightarrow 0$}
\label{sec:lambfinite}

\begin{figure}[htb]
  \begin{center}
    \includegraphics[width=0.95\textwidth]{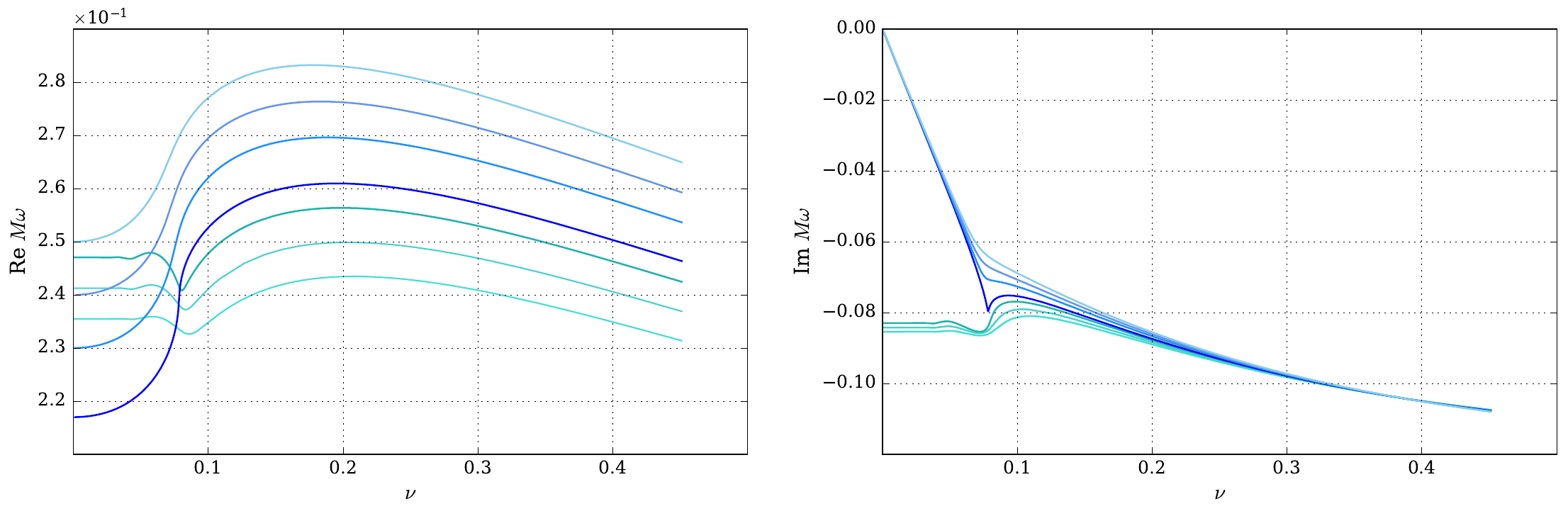}
    \includegraphics[width=0.95\textwidth]{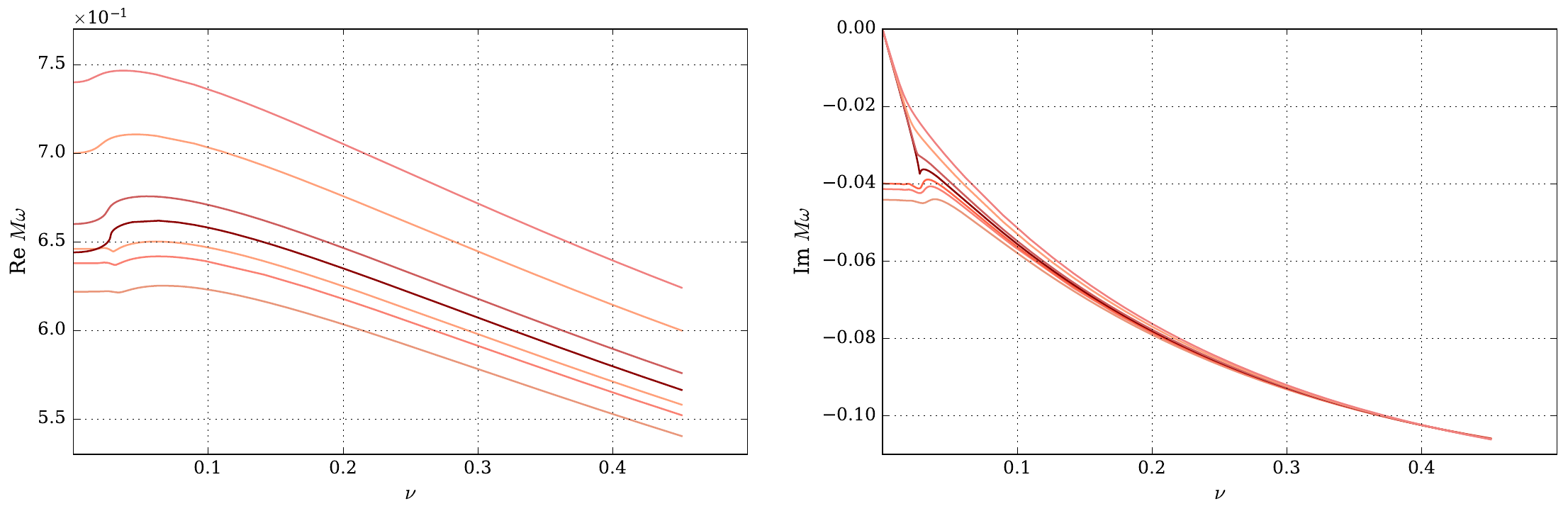}
  \end{center}
  \caption{Damping and non-damping modes for $s=0, l=0$ (top) and
    $s=-\frac{1}{2}$, $\ell =\frac{3}{2}$ (bottom) in near-extremal RN
    black hole. The dark color line identifies the critical values
    $qQ_c\approx 0.216$  ($s=0$, $\ell=0$) and $qQ_c\approx 0.643$
    ($s=-\frac{1}{2}$, $\ell=\frac{1}{2}$).}
  \label{fig:s0l0_nu}
\end{figure}

As the black hole approaches extremality, and $qQ$ is large enough, the
fundamental QNMs frequencies approach the purely real value $M\omega
\rightarrow qQ$. Following \cite{Richartz:2014jla}, we will call these
modes non-damping. Using the isomonodromic method, we can solve
\eqref{eq:radialsystemeqn} numerically and study the behavior of the
spin-$0$ and spin-$\frac{1}{2}$ modes as $\nu\rightarrow 0$. The
result is displayed in Fig. \ref{fig:s0l0_nu}, which can be understood
as a zoomed in version of Fig. \ref{fig:s01over2l01over2} at the
near-extremal region. One notes a drastic
bifurcation at a critical value $qQ_c(s)$, above which the modes
become non-damping. We note that the extremal values for $qQ<qQ_c$
also have a distinct ``almost constant'' behavior for $\nu\rightarrow
0$ in contrast with the $\ell>|s|$ case.

\begin{figure}[htb]
  \begin{center}
    \includegraphics[width=0.95\textwidth]{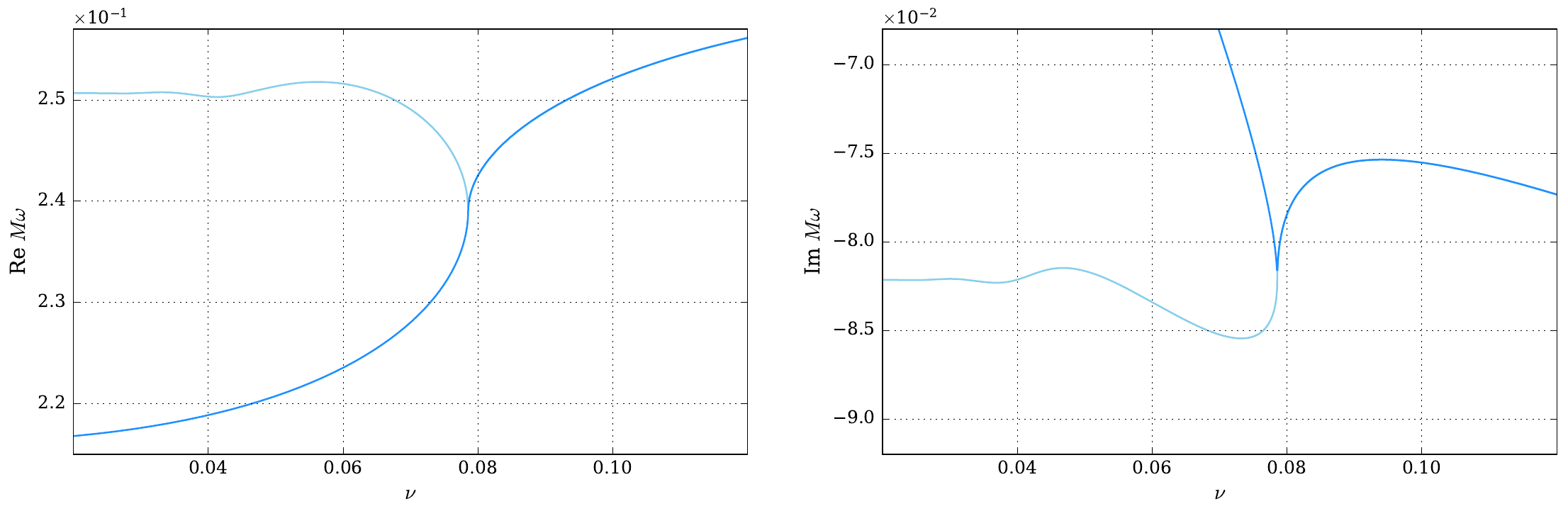}
    \includegraphics[width=0.95\textwidth]{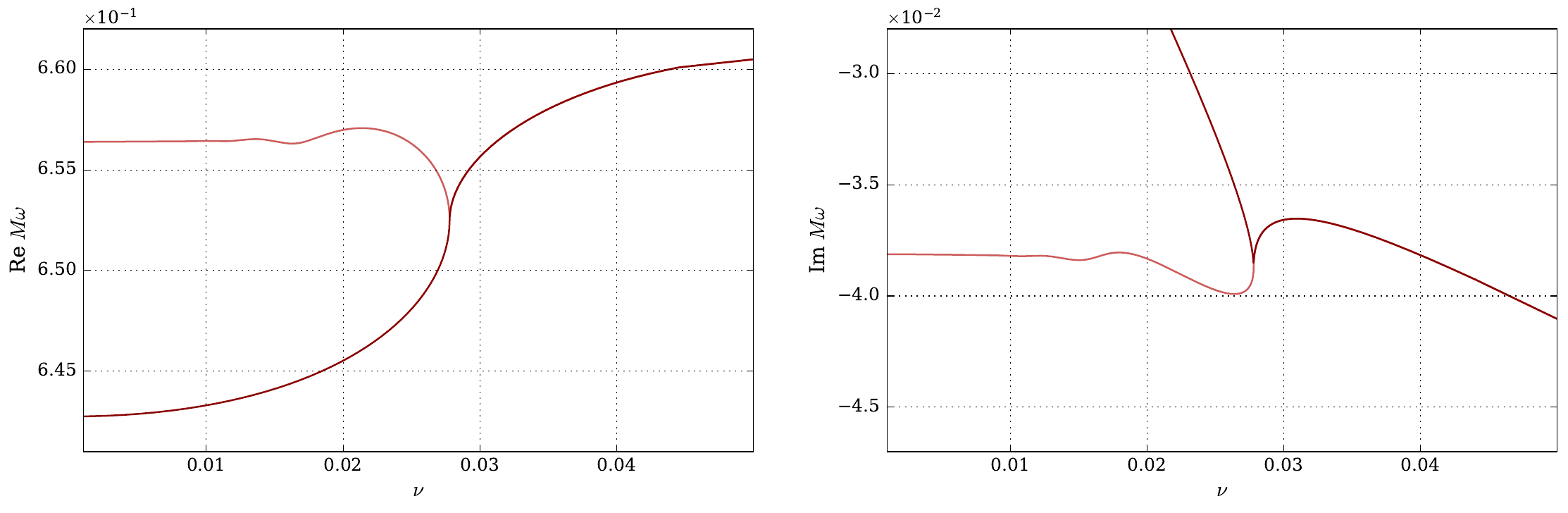}
  \end{center}
  \caption{Non-damping transitions for $\ell = |s|$ scalar (top) and
    spinor (bottom) modes for the RN black hole. The modes represented
    with darker color have coupling slightly above the critical value,
    whereas the lighter color are slightly below. Even with $qQ$
    differing only in the sixth decimal plane, they display radically
    different extremal limits in the region $\nu\leq \nu_c$.} 
  \label{fig:criticalqQ}
\end{figure}

In Fig. \ref{fig:criticalqQ} we zoom into the transition for $s=0$ and
$s=-\frac{1}{2}$. A more careful numerical analysis yields critical values:
\begin{equation}
  qQ_c(s=0) \simeq 0.216228,\qquad
  \text{and }\qquad qQ_c(s=-1/2) \simeq 0.642745,
  \label{eq:criticalqQ}
\end{equation}
with the split at $\nu=\nu_c\simeq 0.078547$ for $s=0$ and
$\nu_c\simeq 0.027797$ for $s=-1/2$, corresponding to $Q/M\simeq
0.996917$ and $Q/M\simeq 0.999614$, respectively.

The non-damping modes also have the property that the associated
single monodromy parameters $\theta_\pm$ in
\eqref{eq:nuparametrization} have finite limits as $\nu\rightarrow
0$. Given that the isomonodromic expansion parameter $z_0$ is
proportional to $\sin\nu$, it is also small in this limit and one can
actually solve the equations \eqref{eq:radialsystemeqn} approximately
by considering the first terms of the expansions involved. For the
calculation of the QNMs frequencies, one can tackle the expansions for
$c_0$ in \eqref{eq:c5expansion} and $\chi_V$ in \eqref{eq:zerochi5} in
terms of $\nu$ directly. We follow the strategy of
\cite{daCunha:2021jkm} and write the schematic expansion for the
frequency and the composite monodromy parameter $\sigma$ as
\begin{equation}
  M\omega = qQ + \beta_1 \nu + \beta_2 \nu^2+\ldots,\qquad
  \sigma = 1 + \alpha_0+\alpha_1 \nu + \alpha_2\nu^2+\ldots,
\end{equation}
where the coefficients $\beta_i$ and $\alpha_i$ may encode
non-analytic corrections in $\nu$, defining $M\omega$ and $\sigma$ as
formal transseries in $\nu$ and $\log\nu$.
The values for $\alpha_i$ and $\beta_i$ can be found recursively from
the expansions \eqref{eq:c5expansion} and \eqref{eq:zerochi5}. For instance,
substitution of $\sigma$ into \eqref{eq:c5expansion} yields
\begin{equation}
\sigma = 1+\alpha_0+ \frac{16qQ
  (2q^2Q^2-s(3+4s)-6{}_{s}\lambda_{\ell})}{\alpha_0(\alpha_0^2-1)}
\beta_1 \nu + \ldots,
\end{equation}
where
\begin{equation}
  {}_{s}\lambda_{\ell} = (s-\ell)(s+\ell+1),\qquad
  \alpha_0 =\sqrt{(1 + 2\ell)^2-4q^2Q^2}.
\end{equation}
With the value of $\sigma$ at hand, we can substitute in
\eqref{eq:zerotau5p} and find an equation for $\beta_1$. We note that
$\chi_V$ in the right-hand side of that equation is analytic in $t_0$
and thus can be written as expansion in $\nu$. The non-analytic terms like
$\log\nu$ arise from the expansion of the gamma functions in
$\Theta_V$ \eqref{eq:theta5}. By considering the lowest-order
contribution in $\nu$, we find, after some algebra,
\begin{equation}
e^{-\frac{i\pi}{2}\alpha_0}\frac{\Gamma(1-\alpha_0)^2 \Gamma(
  \frac{1}{2}(1+\alpha_0)-i\beta_1)\Gamma(\frac{1}{2}
  (1+\alpha_0)-iqQ-s) \Gamma(\frac{1}{2}(1+\alpha_0)-iqQ+
  s)}{\Gamma(1+\alpha_0)^2\Gamma(\frac{1}{2} (1-\alpha_0)-i
  \beta_1)\Gamma (\frac{1}{2} (1-\alpha_0)-iqQ-s) \Gamma (\frac{1}{2}
  (1-\alpha_0)-iqQ+s)}(4\nu qQ)^{\alpha_0} =
1+\mathcal{O}(\nu,\nu\text{log}\nu) .
\label{eq:zerotau5pnu}
\end{equation}
Assuming $\alpha_0$ real and positive, the small $\nu$ regime of
\eqref{eq:zerotau5pnu} simplifies considerably. The $\nu^{\alpha_0}$
term becomes parametrically small, and the equation can be solved
provided the argument of one of the gamma functions becomes very
small, near its pole at vanishing argument. At any rate, this is the
behavior we expect from the numerical analysis above the
critical value $qQ_c<qQ\leq 1$. We therefore find that the first
correction to the non-damping frequency is
\begin{equation}
\beta_1=-\frac{1}{2}i(\alpha_0+1)+i\frac{(-i)^{\alpha_0} \Gamma
  (1-\alpha_0)^2 \Gamma (\frac{1}{2} (\alpha_0+1)-iqQ+s) \Gamma
  (\frac{1}{2} (\alpha_0+1)-iqQ-s)}{\Gamma(-\alpha_0)\Gamma
  (1+\alpha_0)^2 \Gamma (-s-\frac{1}{2}(\alpha_0+1)-iq Q) \Gamma
  (s-\frac{1}{2}(\alpha_0+1)-iqQ)} (4\nu q Q)^{\alpha_0}+\ldots
\end{equation}
again we note that the second term becomes irrelevant for small
$\nu$. Dropping this term and summarizing the result, we find the
frequency to be
\begin{equation}
  M\omega \simeq 
  qQ -\frac{1}{2}i(\alpha_0+1)\nu = qQ
  -\frac{i}{2}(1+\sqrt{(1+2\ell)^2-4q^{2}Q^{2}})\nu. 
  \label{eq:omegaeq}
\end{equation}
These modes, illustrated in Fig. \ref{fig:criticalqQ}, are the
non-damping $qQ>qQ_c$ QNMs' frequencies with imaginary parts that
vanish linearly with $\nu$ as $\nu\rightarrow 0$. For $\alpha_0$ real,
the near-extremal behavior of the real part of $\omega$ is quadratic,
which again corroborates the findings of last Section. Similar remarks
about the non-damping modes have been made by \cite{Richartz:2017qep}
using the CF method, albeit with a slightly different expression for
$\alpha_0$. 

From \eqref{eq:omegaeq} one can also study the regime $qQ\gg \ell$,
which is of interest to field propagation if not for perturbations of
the background. Now the linear correction appears both to the real and
imaginary parts of the frequency, and expanding \eqref{eq:omegaeq} in
this regime yields
\begin{equation}
  \omega \approx \frac{qQ}{M} +2\pi qQ\, T_{BH}-\frac{i}{2}(2\pi)
  T_{BH}
  \label{eq:omegaTBH}
\end{equation}
in terms of the black-hole temperature $T_{BH}=T_+\approx
\frac{\nu}{2\pi M}$. A slightly different version of this result was
presented in \cite{Hod:2012zzb} using the WKB method.

We close by remarking that, despite
the analysis pointing out that the non-damping modes are still
perturbatively stable as $Q\rightarrow M$, the imaginary part becomes
arbitrarily small, so one still has to consider a non-linear analysis
to answer questions about the stability of the RN background. We hope
to return to this issue in future work.

\section{Confluent Limit and Extremal RN Black Hole}
\label{sec:Conflimit_and_RNextremal}

For the modes with $\ell > |s|$ and $\ell = |s|$ with $qQ<qQ_c$, the
QNM frequencies approach the extremal limit still keeping a finite
imaginary part. In such cases, the extremal limit coincides with the
confluent limit of the \eqref{heuneq}, written in terms of the
parameters \eqref{parameters} as
\begin{equation}
\Lambda = \frac{1}{2}(\theta_t+\theta_0), \ \ \ \theta_{\circ}
=\theta_t-\theta_0, \ \ \ u_0 = \Lambda t_0 
, \ \ \ \Lambda \rightarrow \infty,
\label{eq:conflimit}
\end{equation}
leading to the double-confluent Heun equation
\begin{equation}
\frac{d^2
  y}{du^2}+\bigg[\frac{2-\theta_{\circ}}{u}-\frac{u_0}{u^2}\bigg]
\frac{dy}{du}-\bigg[\frac{1}{4} 
+\frac{\theta_{\star}}{2u}+\frac{u_0c_{u_0}-u_0/2}{u^2}\bigg]y(u)=0, 
\label{doubheuneq}
\end{equation}
with two irregular singularities of Poincaré rank 1 at $z=0$ and
$z=\infty$ \cite{NIST:DLMF}.

As seen in \cite{daCunha:2021jkm}, the RH map for the
double-confluent Heun equation can be cast in terms of the Third
Painlevé transcendent. The confluent limit \eqref{eq:conflimit} of
$\tau_V$ yielding its $\tau$ function, and \eqref{RHmap} is replaced by
\begin{equation}
\tau_{III}(\vec{\theta};\sigma,\eta;u_0)=0, \ \ \ \,
z_0\frac{d}{dt}\text{log}
\tau_{III}(\vec{\theta}_{-};\sigma-1,\eta;u_0)-\frac{(\theta_{\circ}-1)^2}{2}=
u_0 c_{u_0}.
\label{eq:RHmapiii}
\end{equation}
Generic expressions for $\tau_{III}$, listed in
\eqref{eq:fredholmIII}, were also given in
\cite{Gamayun:2013auu,daCunha:2021jkm}. We quote the first terms 
\eqref{eq:fredholmIII}
\begin{multline}
\tau_{III}(\theta_\star,\theta_\circ;\sigma,\eta;u) = C_{III}(\vec{\theta},\sigma)
u^{\frac{1}{4}\sigma^2-\frac{1}{8}\theta_\circ^2}e^{\frac{1}{2}u}\\
\times\left(
1-\frac{\sigma-\theta_\circ\theta_\star}{2\sigma^2}u
-\frac{(\sigma+\theta_\circ)(\sigma+\theta_\star)}{4\sigma^2
	(\sigma-1)^2}\kappa_{III}^{-1}u^{1-\sigma} -
\frac{(\sigma-\theta_\circ)(\sigma-\theta_\star)}{4\sigma^2
	(\sigma+1)^2}\kappa_{III}z^{1+\sigma}+{\cal O}(u^2,u^{2\pm
	2\sigma})\right)
\label{eq:tauIIIexpansion}
\end{multline}
where
\begin{equation}
  \kappa_{III}=
  e^{i\pi\eta}\Pi_{III}=e^{i\pi\eta} u^{\sigma}
  \frac{\Gamma(1-\sigma)^2}{\Gamma(1+\sigma)^2}
  \frac{\Gamma(1+\tfrac{1}{2}(\theta_\star+\sigma))}{
    \Gamma(1+\tfrac{1}{2}(\theta_\star-\sigma))}
  \frac{\Gamma(1+\tfrac{1}{2}(\theta_\circ+\sigma))}{
    \Gamma(1+\tfrac{1}{2}(\theta_\circ-\sigma))}.
  \label{eq:kappaIII}
\end{equation}
We list in the Appendix expressions for $\tau_{III}$ in terms of
Fredholm determinants, which we used in our numerical analysis. The
confluent limit of \eqref{eq:c5expansion} can also be taken
\begin{equation}
\begin{aligned}
u_0c_{u_0}=\frac{(\sigma-1)^2-(\theta_\circ-1)^2}{4}
+&\frac{\theta_\circ\theta_\star}{4}\bigg(\frac{1}{\sigma-2}-\frac{1}{\sigma}\bigg)u_0
\\&-\bigg[\frac{\theta_\circ^2\theta_\star^2}{16}\bigg(\frac{1}{(\sigma-2)^3}-\frac{1}{2\sigma^3}\bigg) 
-\frac{\theta_\circ^2+\theta_\star^2-\theta_\circ^2\theta_\star^2}{
	8\sigma(\sigma-2)}
-\frac{(\theta_\circ^2-1)(\theta_\star^2-1)}{8(\sigma+1)(\sigma-3)}      
\bigg]u_0^2+{\cal O}(u_0^3).
\label{eq:c3expansion}
\end{aligned}
\end{equation}
Finally, following the same steps of the Section \eqref{sec:monos}, we
can take the confluent limit \eqref{eq:conflimit} of the condition
\eqref{eq:quantizationV}. We therefore have that the boundary
conditions relating to the QNMs are written in terms of the monodromy
parameters as 
\begin{equation}
e^{i\pi\eta}=e^{-2\pi i\sigma}
\frac{\sin\tfrac{\pi}{2}(\theta_\star+\sigma)}{
	\sin\tfrac{\pi}{2}(\theta_\star-\sigma)}
\frac{\sin\tfrac{\pi}{2}(\theta_\circ+\sigma)}{
	\sin\tfrac{\pi}{2}(\theta_\circ-\sigma)}
\label{eq:quantizationIII}.
\end{equation}

One can now consider the confluent limit of the parameters
\eqref{parameters} of the radial equation \eqref{eq:radeq}.  The
single monodromy parameters are
\begin{equation}
\theta_{\text{ext},\circ} = \theta_{\circ} =  2s
+2i(-qQ+2Q\omega),\qquad
\theta_{\text{ext},\star} = \theta_{\star}=-2s+2i(2Q\omega-qQ)
\label{eq:extremalparameters}
\end{equation}
and accessory parameter and modulus are now obtained by considering
\eqref{modaccASnufunction} in the appropriate confluent limit
\eqref{eq:conflimit}
\begin{equation}
  z_{\mathrm{ext}}c_{z_{\mathrm{ext}}} =
  {}_{s}\lambda_{l,m} +2s-i(1-2s)qQ +
  2(qQ+i(1-2s)) M\omega
  -8(M\omega)^2,
  \qquad
  z_{\mathrm{ext}}= -4M\omega(M\omega-qQ). \\ 
\end{equation}

Formally, the RH map \eqref{eq:RHmapiii} for the double-confluent Heun
equation is written in terms of the extremal parameters
\eqref{eq:extremalparameters} 
\begin{equation}
\tau_{III}(\vec{\theta}_{\mathrm{ext}};\sigma,\eta;z_{\mathrm{ext}})=0,
\qquad
z_{\mathrm{ext}}\frac{d}{dt}\log\tau_{III}(\vec{\theta}_{\mathrm{ext},-};
\sigma-1,\eta;z_{\mathrm{ext}})-
\frac{(\theta_{\mathrm{ext},\circ}-1)^2}{2}=
z_{\mathrm{ext}}c_{z_\mathrm{ext}},
\label{eq:radialextremalsystemeqn}
\end{equation}
with $\vec{\theta}_{\mathrm{ext}} =
\{\theta_{\circ},\theta_{\star}\}$, $\vec{\theta}_{\mathrm{ext},-} =
\{\theta_{\circ}-1,\theta_{\star}-1\}$ and $\eta$ given in
\eqref{eq:quantizationIII}. 

\begin{figure}[htb]
  \begin{center}
    \includegraphics[width=0.95\textwidth]{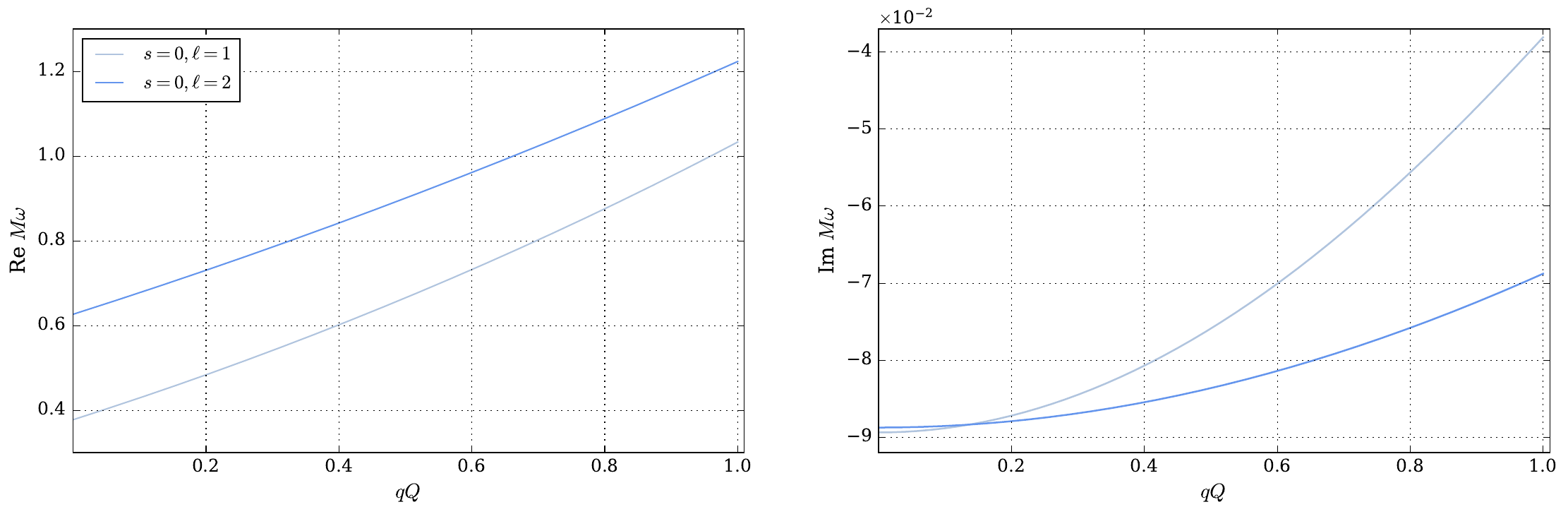}
    \includegraphics[width=0.95\textwidth]{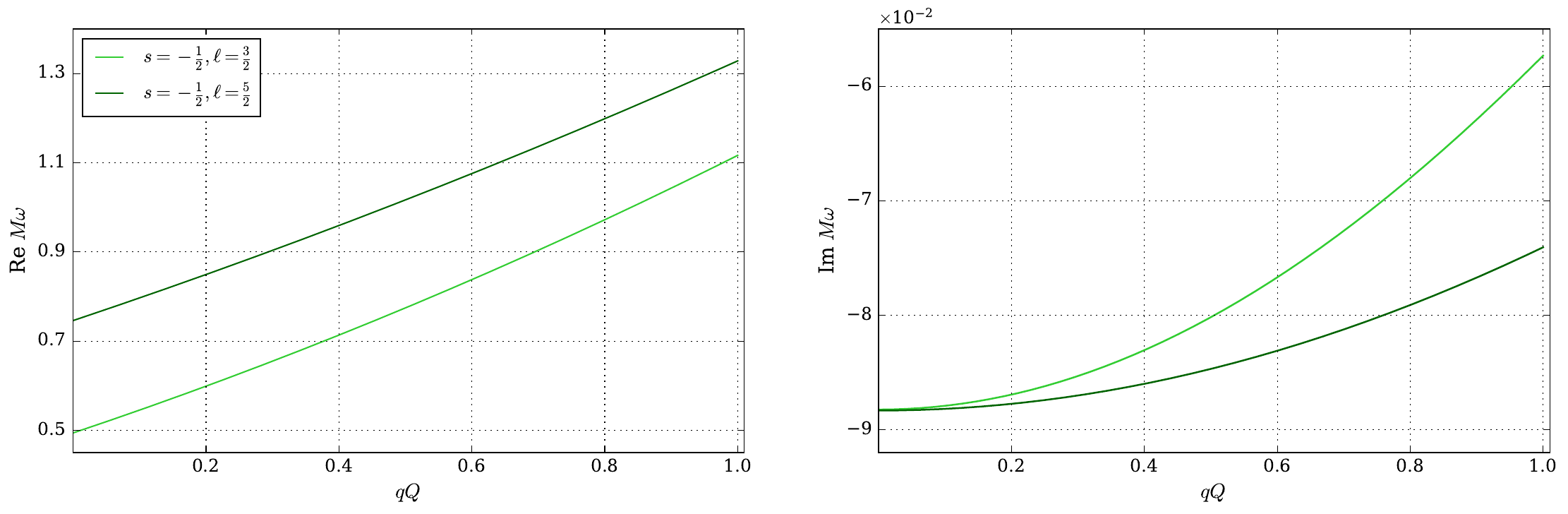}
  \end{center}
  \caption{Fundamental modes for $s=0, l=1,2$ (top) and
    $s=-\frac{1}{2}$, $\ell =\frac{3}{2}, \frac{5}{2}$ (bottom)
    in extremal RN black hole $(Q=M)$ as a function of $qQ$.}
  \label{fig:s01over2_extremal}
\end{figure}

As in the non-extremal case, the resolution of the equations
\eqref{eq:radialextremalsystemeqn} can be done directly given an
implementation of the $\tau_{III}$ function. In
Fig. \ref{fig:s01over2_extremal} we plot the fundamental spin-$0$ and
spin-$\frac{1}{2}$ perturbations for the extremal RN black hole as a
function of $qQ$, for the $\ell>|s|$ case. As expected, modes with
higher values of $\ell$ tend to decay faster, and the difference
increases with $qQ$.

In Tables \eqref{tab:scalar} and \eqref{tab:dirac} we give further
support that the extremal limit of the $\ell>|s|$ modes is actually
smooth, by solving the non-extremal equations
\eqref{eq:radialsystemeqn} at $Q/M$ very close to $1$ and the actual
extremal value computed numerically from
\eqref{eq:radialextremalsystemeqn}. The values for $qQ=0$ in the
scalar case show good accordance with \cite{PhysRevD.53.7033}, whereas
the spinor case agrees with \cite{Richartz:2015saa} for
$qQ<0.1$. Unlike the CF method used in these references, the RH map
allows us to compute the QNMs frequencies for arbitrary values of
$qQ$. 

\begin{figure}[htb]
\begin{center}
  \includegraphics[width=0.95\textwidth]{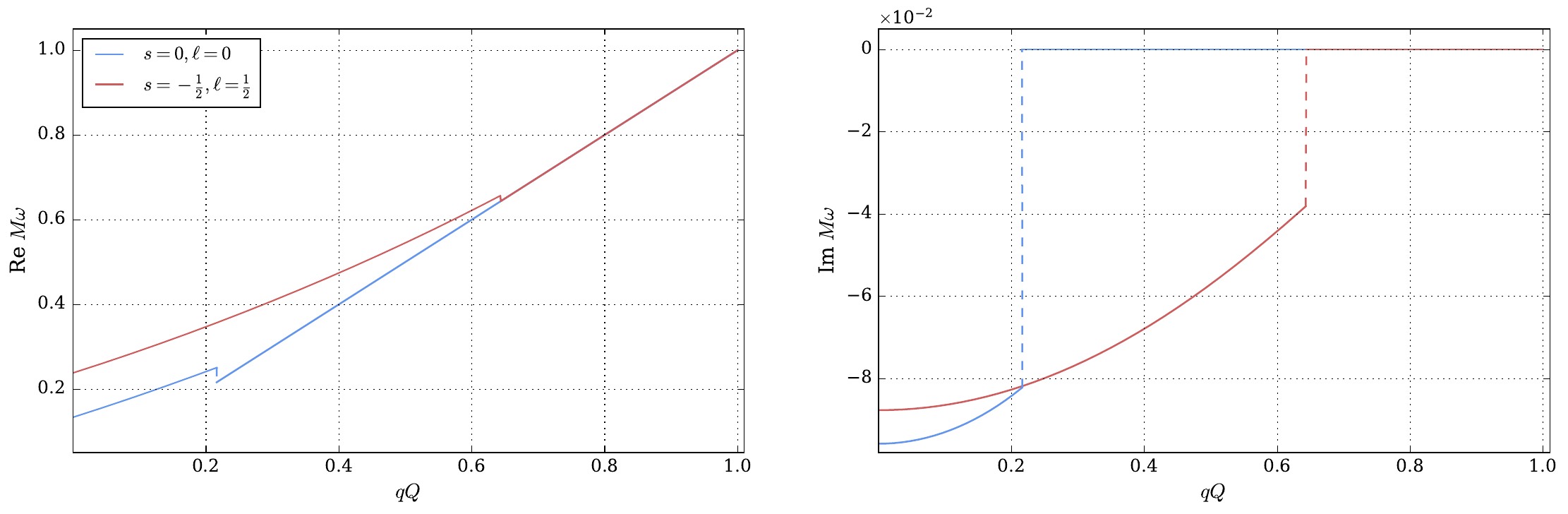}
\end{center}
  \caption{QNMs for $\ell=|s|$ in the scalar (blue) and spinorial
    (red) cases. For the interaction parameter less that the critical
    value of \eqref{eq:criticalqQ}, the values are calculable from the
  $\tau_{III}$ function. Above the critical value, the modes are
  non-damping and the imaginary values of the eigenfrequencies go
  abruptly to zero.} 
  \label{fig:seql_extremal}
\end{figure}

For $\ell = |s|$ the QNM frequencies will be divided into two
different classes, listed in \ref{item:case}.  For the interaction
parameter $qQ$ smaller than the critical value given in
\eqref{eq:criticalqQ}, the behavior of the QNM is similar to the $\ell
>|s|$ case in which there will be a non-zero damping factor in the
extremal limit. The values for the frequencies at extremality can also
be computed from the equations \eqref{eq:radialextremalsystemeqn}. For
$qQ$ above the critical value, however, the mode ``decouples'' from
the double-confluent equations, and the frequency becomes non-damped
with value $qQ/M$. The behavior is illustrated in Fig.
\ref{fig:seql_extremal}. Because of the non-damping modes, the question of
stability of the extremal RN black hole under scalar and spinorial
perturbations can only be settled by considering higher order terms in
the Maxwell-Einstein equations.

\setlength\tabcolsep{1.5mm}
{\renewcommand{\arraystretch}{1.2}
  \begin{table}[htb]
    \begin{tabular}{|c|c|c|c|c|}
      \hline
      qQ& \multicolumn{2}{c|}{$\ell=1$} & \multicolumn{2}{c|}{$\ell=2$}  \\ \hline
      &  $Q/M=0.999999$ &  $Q/M=1$  & $Q/M=0.999999$ & $Q/M=1$    \\
      \hline
      0.0 & $0.3776416 - 0.0893845i$  & $0.3776418 - 0.0893843i$  &
                                                                    $0.6265722
                                                                    -
                                                                    0.0887485i$
                              &   $0.6265727 - 0.0887483i$    \\
      \hline
      0.1 & $0.4291343 - 0.0888385i$  & $0.4291346 - 0.0888382i$  &
                                                                    $0.6775330 - 0.0885422i$   &   $0.6775336 - 0.0885419i$   \\ \hline
      0.2 &  $0.4836193 - 0.0872035i$ & $0.4836196 - 0.0872031i$  &
                                                                    $0.7304163 - 0.0879239i$   &    $0.7304170 - 0.0879236i$       \\ \hline
      0.3 & $0.5411167 - 0.0844886i$  & $0.5411172 - 0.0844882i$  &
                                                                    $0.7852239
                                                                    -
                                                                    0.0868961i$
                              &     $0.7852248 - 0.0868957i$      \\
      \hline
      0.4 & $0.6016610 - 0.0807101i$  & $0.6016616 - 0.0807095i$  &
                                                                    $0.8419591 - 0.0854624i$   &     $0.8419601 - 0.0854620i$      \\ \hline
      0.5 & $0.6653021 - 0.0758925i$  & $0.6653028 - 0.0758917i$  &
                                                                    $0.9006264
                                                                    -
                                                                    0.0836285i$
                              &     $0.9006276 - 0.0836280i$      \\
      \hline
      0.6 & $0.7321070 - 0.0700711i$  & $0.7321078 - 0.0700701i$  &
                                                                    $0.9612317 - 0.0814015i$   &     $0.9612330 - 0.0814008i$      \\ \hline
      0.7 & $0.8021625 - 0.0632953i$  & $0.8021634 - 0.0632939i$  &
                                                                    $1.0237821 - 0.0787904i$   &     $1.0237836 - 0.0787896i$      \\ \hline
      0.8 & $0.8755790 - 0.0556340i$  & $0.8755799 - 0.0556320i$  &
                                                                    $1.0882862
                                                                    -
                                                                    0.0758061i$
                              &    $1.0882879 - 0.0758052i$       \\
      \hline
      0.9 & $0.9524949 - 0.0471845i$  & $0.9524959 - 0.0471819i$  &
                                                                    $1.1547541
                                                                    -
                                                                    0.0724620i$
                              &    $1.1547561 - 0.0724608i$       \\
      \hline
      1.0 & $1.0330846 - 0.0380869i$  & $1.0330854 - 0.0380833i$  &
                                                                    $1.2231973
                                                                    -
                                                                    0.0687733i$
                              &     $1.2231995 - 0.0687720i$      \\
      \hline
    \end{tabular}
    \caption{A comparison between the fundamental modes for scalar
      $s=0$ perturbations of the RN black hole computed from
      \eqref{eq:radialsystemeqn} with $Q/M$ very close to $1$, and
      those computed in the extremal case using
      \eqref{eq:radialextremalsystemeqn} for various values of the
      interaction parameter $qQ$.}
   \label{tab:scalar}
\end{table}
\begin{table}[htb]
  \begin{tabular}{|c|c|c|c|c|}
    \hline
    qQ& \multicolumn{2}{c|}{$\ell=3/2$} &
                                          \multicolumn{2}{c|}{$\ell=5/2$}  \\ \hline
    &  $Q/M=0.999999$ & $Q/M=1$  & $Q/M=0.999999$    &
                                                                  $Q/M=1$    \\ \hline
    0.0 & $0.4941127 - 0.0882400i$  &  $0.4941131 - 0.0882398i$  &
                                                                   $0.7460834
                                                                   -
                                                                   0.0883267i$
                      &     $0.7460841 - 0.0883265i$     \\ \hline
    0.1 & $0.5453239 - 0.0879134i$  & $0.5453244 - 0.0879132i$  &
                                                                  $0.7969051
                                                                  -
                                                                  0.0881804i$
                      &    $0.7969059 - 0.0881802i$       \\ \hline
    0.2 & $0.5989590 - 0.0869355i$  & $0.5989596 - 0.0869352i$  &
                                                                  $0.8493706
                                                                  -
                                                                  0.0877419i$
                      &     $0.8493715 - 0.0877416i$      \\ \hline
    0.3 &  $0.6550227 - 0.0853120i$  & $0.6550234 - 0.0853119i$  &
                                                                   $0.9034806
                                                                   -
                                                                   0.0870124i$
                      &   $0.9034816 - 0.0870120i$        \\ \hline
    0.4 & $0.7135223 - 0.0830539i$  & $0.7135231 - 0.0830534i$  &
                                                                  $0.9592359
                                                                  -
                                                                  0.0859942i$
                      &    $0.9592370 - 0.0859938i$       \\ \hline
    0.5 & $0.7744686 - 0.0801747i$  & $0.7744696 - 0.0801740i$  &
                                                                  $1.0166382
                                                                  -
                                                                  0.0846905i$
                      &     $1.0166394 - 0.0846901i$      \\ \hline
    0.6 &  $0.8378753 - 0.0766939i$  & $0.8378764 - 0.0766930i$  &
                                                                   $1.0756891
                                                                   -
                                                                   0.0831055i$
                      &    $1.0756906 - 0.0831049i$       \\ \hline
    0.7 & $0.9037594 - 0.0726358i$  & $0.9037607 - 0.0726347i$  &
                                                                  $1.1363911
                                                                  -
                                                                  0.0812441i$
                      &    $1.1363927 - 0.0812434i$       \\ \hline
    0.8 & $0.9721413 - 0.0680306i$  & $0.9721429 - 0.0680293i$  &
                                                                  $1.1987468
                                                                  -
                                                                  0.0791125i$
                      &    $1.1987486 - 0.0791117i$       \\ \hline
    0.9 & $1.0430450 - 0.0629152i$  & $1.0430468 - 0.0629135i$  &
                                                                  $1.2627593
                                                                  -
                                                                  0.0767178i$
                      &     $1.2627613 - 0.0767169i$      \\ \hline
    1.0 & $1.1164980 - 0.0573342i$  & $1.1165001 - 0.0573321i$  &
                                                                  $1.3284321 - 0.0740684i$   &     $1.3284344 - 0.0740674i$      \\ \hline
  \end{tabular}
  \caption{The comparison between fundamental modes for spinorial $s=-1/2$
    perturbations of the RN black hole obtained from near-extremal
    \eqref{eq:radialsystemeqn} and extremal
    \eqref{eq:radialextremalsystemeqn}
    as a function of  $qQ$.}
  \label{tab:dirac}
\end{table}
}

\section{Discussion}
\label{sec:discussion}

In this paper the authors used the isomonodromic method to study
quasinormal modes frequencies for scalar and spinorial perturbations
of the Reissner-Nordström black hole. The map was derived by the
authors in previous work \cite{CarneirodaCunha:2019tia}, constructing
monodromy parameters for generic solutions of the confluent Heun
equation and making use of the expansions of the fifth and third
Painlevé transcendents by
\cite{Gamayun:2013auu,Lisovyy:2018mnj,daCunha:2021jkm}, itself sparked
by the deep relation between instanton counting and Liouville
conformal blocks \cite{Alday:2009aq,Alba:2010qc}. Although we have
made use of the ``$c=1$'' version of the relation -- or rather, the
Riemann-Hilbert problem formulation of the relation, there is growing
interest in the semiclassical ``Nekrasov-Shatashvilii'' version of the
relation \cite{Jeong:2020uxz,Bershtein:2021uts}, which also proposes
to compute accessory parameters for Fuchsian differential equations,
and their confluent limits, given monodromy data. For the practical
purpose of solving boundary problems like finding QNM frequencies,
expressions like \eqref{RHmap} and \eqref{eq:RHmapiii} provide not only a
formal solution to the system, but also a means of calculation which
is unencumbered by the shortcomings of the usual CF and WKB methods
\cite{Aminov:2020yma}. In \cite{Bonelli:2021uvf}, the authors gave a
comprehensive list of properties of the Kerr background like greybody
factors, QNMs and Love numbers in terms of monodromy parameters.

For the RN black hole, the analysis is comparatively simpler, since
the angular problem can be solved exactly in terms of spin-weighted
spherical harmonics. We have found that the behavior of the QNMs
parallels those found for Kerr in \cite{daCunha:2021jkm}, along with
the modes which decay even in the extremal limit, there are
non-damping modes for $\ell=|s|$ whose relaxation time diverges with
the inverse temperature as one approaches $Q=M$. These non-damping
modes parallel the co-rotating QNMs in Kerr geometry, and we have
found the transition to happen above a critical value for the interaction
parameter \eqref{eq:criticalqQ}. This parameter plays a role for RN
similar to the angular momentum to Kerr, and, unlike $m$, it
can be varied continuously. We have found the critical value $qQ_c$ to
be a bifurcation point in the sense that for small enough extremality
parameter $\nu$, the behavior of the QNMs changes abruptly as the
interaction parameter crosses $qQ_c$. The actual mechanism for this
transition certainly needs clarification, and we hope to return to
this issue in the future. Here we have kept our analysis to
scalar and spinor modes, but, given the wide scope of the RH map, we
have every reason to expect that the method will work for higher spin
perturbations as well. 

We hope that the method developed in the aforementioned papers and
applied here proves to be an invaluable tool in the calculation of
black hole perturbations, both in the asymptotically flat case and
otherwise
\cite{Novaes:2018fry,Barragan-Amado:2018pxh,Amado:2020zsr}. The
particular case of Reissner-Nordström black holes in spaces with
positive cosmological constant has received a great deal of attention
following the argument for the violation of the strong cosmic
censorship conjecture in \cite{Hollands:2019whz}. The presence of the
cosmological constant requires the Riemann-Hilbert map associated to the
non-confluent Heun equation, which in turn is related to the Painlevé
VI $\tau$ function \cite{Novaes:2014lha,Novaes:2018fry}, and the
Reissner-Nordström case will be dealt with in future work. With
the comparatively nice properties of the isomonodromic
$\tau$-functions involved, one can perhaps utilize it to even shed
light on the full, nonlinear, stability issues \cite{Green:2019nam}. 

\section{Acknowledgements}

The authors thank A. Grassi, A. Tanzini and S. Hod for comments and
suggestions. JPC acknowledges partial support from CNPq.

\appendix
\section{$\tau_{V}$ and $\tau_{III}$ function and Quantization condition }
\label{sec:tools}

The Fredholm determinant expression for $\tau_V$ is given by
\begin{equation}
  \tau_V(\vec{\theta};\sigma,\eta;t)=
  t^{\tfrac{1}{4}(\sigma^2-\theta_0^2-\theta_t^2)}
  e^{\tfrac{1}{2}\theta_t t}
  \det(\mathbbold{1}-\mathsf{A}
  \kappa_V^{\tfrac{1}{2}\sigma_3}t^{\tfrac{1}{2}\sigma\sigma_3}
  \mathsf{D}_c(t)
  \kappa_V^{-\tfrac{1}{2}\sigma_3}t^{-\tfrac{1}{2}\sigma\sigma_3})
  \label{eq:fredholmV}
\end{equation}
where $\mathsf{A}$ and $\mathsf{D}_c$ are operators acting on pair of
analytic functions defined on a circle ${\cal C}$ of radius $R<1$,
\begin{equation}
  (\mathsf{A}g)(z)=\oint_{\cal C} \frac{dz'}{2\pi i}A(z,z')g(z'),\qquad
  (\mathsf{D}_cg)(z)=\oint_{\cal C} \frac{dz'}{2\pi i}D_c(z,z')g(z'),\qquad
  g(z')=\begin{pmatrix}
    f_+(z) \\
    f_-(z)
  \end{pmatrix}
  \label{eq:fredholmad}
\end{equation}
with kernels given explicitly for $|t|<R$, by
\begin{equation}
  \begin{gathered}
    A(z,z')=\frac{\Psi^{-1}(\sigma,\theta_t,\theta_0;z')
      \Psi(\sigma,\theta_t,\theta_0;
      z)-\mathbbold{1}}{z-z'},\\ 
    D_c(z,z')=\frac{\mathbbold{1}-\Psi_c^{-1}(-\sigma,\theta_\star;t/z')
      \Psi_c(-\sigma,\theta_\star;t/z)}{z-z'},
  \end{gathered}
\end{equation}
where the parametrices $\Psi$ and $\Psi_c$ are matrices whose entries
are given by
\begin{equation}
  \displaystyle
    \Psi(\sigma,\theta_t,\theta_0;z) = \begin{pmatrix}
      \phi(\sigma,\theta_t,\theta_0;z) &
      \chi(-\sigma,\theta_t,\theta_0;z) \\
      \chi(\sigma,\theta_t,\theta_0;z) &
      \phi(-\sigma,\theta_t,\theta_0,z)
    \end{pmatrix},
\end{equation}
with $\phi$ and $\chi$ in terms of Gauss' hypergeometric series
\begin{equation}
  \displaystyle 
  \begin{gathered}
    \phi(\sigma,\theta_t,\theta_0;z) = {_2F_1}(
    \tfrac{1}{2}(\sigma-\theta_t+\theta_0),\tfrac{1}{2}(\sigma-\theta_t-\theta_0);
    \sigma;z) \\
    \chi(\sigma,\theta_t,\theta_0;z) =
    \frac{\theta_0^2-(\sigma-\theta_t)^2}{4\sigma(1+\sigma)}
      z\,{_2F_1}(
      1+\tfrac{1}{2}(\sigma-\theta_t+\theta_0),
      1+\tfrac{1}{2}(\sigma-\theta_t-\theta_0);
      2+\sigma;z)
  \end{gathered}
\end{equation}
and 
\begin{gather}
  \label{eq:confluentparametrix}
  \Psi_c(-\sigma,\theta_\star;t/z)=
  \begin{pmatrix}
    \phi_c(-\sigma,\theta_\star;t/z) &
    \chi_c(-\sigma,\theta_\star; t/z) \\
    \chi_c(\sigma,\theta_\star; t/z) &
    \phi_c(\sigma,\theta_\star; t/z)
  \end{pmatrix},
  \nonumber \\
  \phi_c(\pm\sigma,\theta_\star;t/z) = 
  {_1F_1}(\tfrac{-\theta_\star\pm\sigma}{2};\pm\sigma;-t/z), \\
  \chi_c(\pm\sigma,\theta_\star;t/z) =
  \pm\frac{-\theta_\star\pm\sigma}{2\sigma(1\pm\sigma)}\,
  \frac{t}{z}
  \,{_1F_1}(1+\tfrac{-\theta_\star\pm\sigma}{2},2\pm\sigma;-t/z),
\end{gather}
where  ${_1F_1}$ the confluent (Kummer's) hypergeometric
series. Finally, the $\kappa_V$ parameter is defined in terms of
monodromy data by \eqref{eq:kappaV}.

The version of the third Painlevé transcendent $\tau$-function used
here was derived in \cite{daCunha:2021jkm}. The expression is formally
similar to \eqref{eq:fredholmV}
\begin{equation}
  \tau_{III}(\theta_\star,\theta_\circ;\sigma,\eta;u) =
  u^{\frac{1}{4}\sigma^2-\frac{1}{8}\theta_\circ^2}
  e^{\frac{1}{2}u}\det(
  \mathbbold{1}-\mathsf{A}_c\kappa_{III}^{\frac{1}{2}\sigma_3}
  u^{\frac{1}{2}\sigma\sigma_3}\mathsf{D}_c(u)
  \kappa_{III}^{-\frac{1}{2}\sigma_3}u^{-\frac{1}{2}\sigma\sigma_3}).
  \label{eq:fredholmIII}
\end{equation}
where $\mathsf{D}_c$ is as above and the kernel of $\mathsf{A}_c$ is
now also given by confluent hypergeometric functions
\begin{equation}
  A_c(z,z')=\frac{\Psi_c^{-1}(\sigma,\theta_\circ;z')
    \Psi_c(\sigma,\theta_\circ;z)-\mathbbold{1}}{z-z'},
\end{equation}
with $\Psi_c$ as in \eqref{eq:confluentparametrix} and $\kappa_{III}$
given in terms of monodromy data by \eqref{eq:kappaIII}.

If one starts from \eqref{eq:fredholmV} and \eqref{eq:fredholmIII}, by
expanding the determinants to first order in $t$ and $u$ one recovers
the first terms in the expansions given by \eqref{eq:expansiontauV}
and \eqref{eq:tauIIIexpansion}, respectively. In order to make use of
these expressions numerically, we wrote the matrix elements of the
operators in a Fourier basis -- on the circle ${\cal C}$ --
effectively reducing the Fredholm operator to a finite matrix after
truncating the basis to order $N_f$. The interested reader can find
the details in \cite{daCunha:2021jkm} or in \cite{Github}, where an
implementation in Julia is available.

\subsection{Quantization Condition} 
\label{sec:quantcond}

To derive the quantization condition one uses the following set of
local solutions of \eqref{heuneq} near $z=t_0$ and $z=\infty$,
\begin{gather}
y_{t_0,+}(z)=(z-t_0)^{\theta_{t}}(1+{\cal O}(z-t_0)),\qquad
y_{t_0,-}(z)=(z-t_0)^{0}(1+{\cal O}(z-t_0)),\\
y_{\infty,+}(z)=e^{z/2}z^{-\theta_\star/2}(1+{\cal O}(1/z)),\qquad
y_{\infty,-}(z)=e^{-z/2}z^{\theta_\star/2}(1+{\cal O}(1/z)),
\end{gather}
where $y_{t,\pm}$ form a basis of solutions near
$z=t_0$, while $y_{\infty,\pm}$ represent a basis near of $z=\infty$.

After some algebraic manipulation of the monodromy matrices, the
connection matrix $\mathsf{C}_t$ between these local solutions are
written in the following form 
\begin{equation}
\begin{pmatrix}
\rho_{\infty}y_{\infty,+}(z) \\
\tilde{\rho}_{\infty}y_{\infty,-}(z)
\end{pmatrix}
=\mathsf{C}_{t}
\begin{pmatrix}
\rho_{t_0}y_{t_0,+}(z)\\
\tilde{\rho}_{t_0}y_{t_0,-}(z)
\end{pmatrix}
=\begin{pmatrix}
e^{-\tfrac{i\pi}{2}\eta}\zeta'_{t_0}-e^{\tfrac{i\pi}{2}\eta}\zeta_{t_0}
& 
-e^{-\tfrac{i\pi}{2}\eta}\zeta_{\infty}\zeta'_{t_0} +
e^{\tfrac{i\pi}{2}\eta}\zeta'_{\infty}\zeta_{t_0} \\
e^{-\tfrac{i\pi}{2}\eta}-e^{\tfrac{i\pi}{2}\eta} &
-e^{-\tfrac{i\pi}{2}\eta}\zeta_{\infty}+e^{\tfrac{i\pi}{2}\eta}\zeta'_{\infty}
\end{pmatrix}
\begin{pmatrix}
\rho_{t_0}y_{t_0,+}(z)\\
\tilde{\rho}_{t_0}y_{t_0,-}(z)
\end{pmatrix}
\end{equation}
with
\begin{equation}
\begin{gathered}
\zeta_{\infty} =  e^{\frac{i\pi}{2}\sigma}\sin\tfrac{\pi}{2}
(\theta_\star+\sigma),
\quad\quad
\zeta'_{\infty}= e^{-\frac{i\pi}{2}\sigma}\sin\tfrac{\pi}{2}
(\theta_\star-\sigma),\\
\zeta_{t_0}=\sin\tfrac{\pi}{2}(\theta_t+\theta_0-\sigma)
\sin\tfrac{\pi}{2}(\theta_t-\theta_0-\sigma),\quad\quad
\zeta'_{t_0}=\sin\tfrac{\pi}{2}(\theta_t+\theta_0+\sigma)
\sin\tfrac{\pi}{2}(\theta_t-\theta_0+\sigma),
\end{gathered}
\end{equation}
and $\rho_t,\tilde{\rho}_t,\rho_\infty,\tilde{\rho}_\infty$ are (in
principle arbitrary) normalization constants. In the calculation of
quasinormal modes, we should constraint $\mathsf{C}_t$ to be lower
triangular. We then have an expression for $\eta$  
\begin{equation}
e^{i\pi\eta}=\frac{\zeta_{\infty}\zeta'_{t_0}}{\zeta'_{\infty}\zeta_{t_0}},
\end{equation}
which, when written in terms of the monodromy parameters, gives
\eqref{eq:quantizationV}.

%\bibliography{monodromias}
%\bibliographystyle{JHEP}

\providecommand{\href}[2]{#2}\begingroup\raggedright\endgroup

\end{document}